\documentclass[apj]{emulateapj}

\newcommand{\hi}          {\mbox{\rm \ion{H}{1}}}

\newcommand{\surfb}       {mag~arcsec$^{-2}$}
\newcommand{\kms}         {km~s$^{-1}$}

\newcommand{\msun}        {M$_{\odot}$}
\newcommand{\lsun}        {L$_{\odot}$}
\newcommand{\mlv}         {M$_{\odot}/$L$_{\odot, V}$}
\newcommand{\simgtr}    {\; \raisebox{-.2ex}{$\stackrel{>}{\mbox{\tiny
$\sim$}}$} \;}
\newcommand{\simless}   {\; \raisebox{-.2ex}{$\stackrel{<}{\mbox{\tiny
$\sim$}}$} \;}

\shorttitle{Kinematics of Ultra-Faint Milky Way Satellites}
\shortauthors{Simon \& Geha}
\slugcomment{Accepted for publication in ApJ}

\begin{document}

\title{The Kinematics of the Ultra-Faint Milky Way Satellites: Solving
  the Missing Satellite Problem}

\author{Joshua D. Simon}

\affil{Department of Astronomy, 
       California Institute of Technology, 1200 E. California Blvd.,
       MS 105-24, Pasadena, CA  91125}

\medskip
\and
\vspace{-0.3cm}

\author{Marla Geha}

\affil{National Research Council of Canada, Herzberg Institute of
  Astrophysics, 5071 West Saanich Road, Victoria, BC V9E 2E7, Canada}

\email{jsimon@astro.caltech.edu}
\email{marla.geha@nrc-cnrc.gc.ca}

\begin{abstract}
We present Keck/DEIMOS spectroscopy of stars in 8 of the newly
discovered ultra-faint dwarf galaxies around the Milky Way.  We
measure the velocity dispersions of Canes Venatici~I, Canes
Venatici~II, Coma Berenices, Hercules, Leo~IV, Leo~T, Ursa Major~I,
and Ursa Major~II from the velocities of $18-214$ stars in each galaxy
and find dispersions ranging from 3.3 to 7.6~\kms.  The 6 galaxies
with absolute magnitudes $M_{V} < -4$ are highly dark
matter-dominated, with mass-to-light ratios approaching 1000~\mlv.
For the fainter galaxies, Ursa Major II and (to a lesser extent) Coma
Berenices, we find tentative evidence for tidal disruption, which for
UMa~II is strongly supported by previous studies.  If these 2 galaxies
are also dark matter-dominated, they have extremely large
mass-to-light ratios.  The measured velocity dispersions of the
ultra-faint dwarf galaxies are correlated with their luminosities,
indicating that a minimum mass for luminous galactic systems may not
yet have been reached.  We also measure the metallicities of the
observed stars and find that several of the new dwarfs have mean
metallicities as low as [Fe/H] =$-2.3$; these galaxies represent some
of the most metal-poor known stellar systems.  The 6 brightest of the
ultra-faint dwarfs extend the luminosity-metallicity relationship
followed by more luminous dwarfs by $\sim2$ orders of magnitude in
luminosity.  We detect metallicity spreads of up to 0.5~dex in several
objects, suggesting multiple star formation epochs.  UMa~II and Com,
despite their exceptionally low luminosities, have higher
metallicities that suggest they may once have been much more massive.
Having established the masses of the ultra-faint dwarfs, we re-examine
the missing satellite problem.  After correcting for the sky coverage
of the Sloan Digital Sky Survey, we find that the ultra-faint dwarfs
substantially alleviate the discrepancy between the predicted and
observed numbers of satellites around the Milky Way, but there are
still a factor of $\sim4$ too few dwarf galaxies over a significant
range of masses.  We show that if galaxy formation in low-mass dark
matter halos is strongly suppressed after reionization, the simulated
circular velocity function of CDM subhalos can be brought into
approximate agreement with the observed circular velocity function of
Milky Way satellite galaxies.
\end{abstract}


\section{INTRODUCTION}
\label{intro}

The Cold Dark Matter (CDM) cosmological model predicts that massive
galaxies such as the Milky Way should be surrounded by large numbers
of dark matter-dominated satellite halos.  The relatively modest
populations of observed dwarf galaxies orbiting the Milky Way and
Andromeda, however, seem to conflict with this prediction
\citep*{kgg93,klypin99,moore99}.  This apparent disagreement between
the expected and observed numbers of dwarf galaxies has become widely
known as the ``substructure'' or ``missing dwarf'' problem.

Proposed solutions to the substructure problem can be broadly divided
into two categories: cosmological and astrophysical.  Cosmological
solutions include modifying the power spectrum at small scales
\citep{kl99,zb03} and changing the properties of the dark matter
particles, such as by making them warm \citep*{colin00,bode01} or
invoking a late decay from a non-relativistic particle \citep{skb07}.
Astrophysical solutions are more prosaic, but perhaps easier to
constrain observationally.  Some of the most popular astrophysical
solutions include the hypothesis that reionization could suppress the
formation of dwarf galaxies by preventing low-mass dark matter halos
from acquiring enough gas to form stars after $z\sim10$
\citep*[e.g.,][]{bkw00,somerville02,benson02,rg05,moore06} and the
possibility that the dwarf galaxies we observe today were once much
more massive objects that have been reduced to their present
appearance by dramatic tidal stripping \citep*{mayer01a, mayer01b,
  kravtsov04}.  Despite a wealth of ideas about how to solve the
missing dwarf problem, distinguishing between the various proposals
has proved to be difficult, and making sense of the tremendous variety
of masses, luminosities, mass-to-light ratios, gas fractions, and star
formation histories among observed dwarf galaxies remains a challenge.

Our understanding of the missing satellite problem and the evolution
of dwarf galaxies is being rapidly revised by the discovery of a large
population of new, very faint Local Group dwarfs in the Sloan Digital
Sky Survey \citep[SDSS;][]{sdss} and other wide-field imaging surveys.
In the past 3 years, at least 20 of these galaxies have been
identified, nearly doubling the previously known population.  The new
dwarfs include 8 additional Milky Way dwarf spheroidals
\citep{willman05a,zucker06a,belok06,zucker06b,grillmair06,belok07,sh06}
and 1 dwarf irregular \citep{irwin07}, 8 new dwarf spheroidals around
Andromeda \citep{zucker04,zucker06c,martin06,majewski07,ibata07}, and
3 further new Milky Way satellites that lie in the uncertain parameter
space between dwarf galaxies and globular clusters \citep*{willman05b,
  belok07, walsh07}.  Nearly all of these objects have both surface
brightnesses and luminosities that are significantly lower than those
of any previously-known galaxies.

Properly placing these new discoveries within the framework of CDM and
the missing satellite problem requires measurements of their internal
kinematics, in order to determine whether the ultra-faint dwarfs are
gravitationally bound, dark matter-dominated galaxies, or
tidally-disrupted systems.  Only 5 of these objects (Ursa Major~I,
Andromeda~IX, Bo{\" o}tes, Canes Venatici~I, and Andromeda~XIV) have
published stellar kinematics measurements, and for 2 of the 3
ultra-faint Milky Way dwarfs that have been studied already only a
handful of stars were observed
\citep{kleyna05,chapman05,munoz06,ibata06,majewski07}.  In this paper,
we present new stellar velocity measurements of larger samples of
stars in 8 of the 12 new Milky Way satellites (see Table
\ref{targettable}).  Including the other published and in preparation
studies that we are aware of, the only known Milky Way satellites that
remain unobserved are Segue 1 and Bo{\" o}tes~II.

\begin{deluxetable*}{l c c c c c l}
\tablewidth{0pt}
\tablecolumns{7}
\tablecaption{Observing Targets}
\tablehead{
\colhead{Galaxy} & \colhead{$\alpha$ (J2000.0)}  & 
\colhead{$\delta$ (J2000.0)} & \colhead{$M_{V}$} &
\colhead{$\mu_{V}$\tablenotemark{a}} & \colhead{Distance\tablenotemark{b}} & 
\colhead{References} \\
\colhead{} & \colhead{(h$\,$ $\,$m$\,$ $\,$s)} &
\colhead{($^\circ\,$ $\,'\,$ $\,''$)} & \colhead{} &
\colhead{(\surfb)} & \colhead{(kpc)} & \colhead{}}

\startdata 
 Ursa Major II     & 08 51 30.00 & \phs63 07 48.0 & $-3.8$ & 28.8 & \phn32 & \phm{(4}(1),(2) \\
 Leo T             & 09 34 53.40 & \phs17 03 05.0 & $-7.1$ & 26.9 & 417 & \phm{(4),}(3)     \\
 Ursa Major I      & 10 34 52.80 & \phs51 55 12.0 & $-5.6$ & 28.9 & 106 & (4),(5),(7) \\
 Leo IV            & 11 32 57.00 & $-$00 32 00.0  & $-5.1$ & 28.3 & 158 & \phm{(4),}(2)     \\
 Coma Berenices    & 12 26 59.00 & \phs23 54 15.0 & $-3.7$ & 27.4 & \phn44 & \phm{(4),}(2)  \\
 Canes Venatici II & 12 57 10.00 & \phs34 19 15.0 & $-4.8$ & 27.2 & 151 & \phm{(4),}(2)     \\
 Canes Venatici I  & 13 28 03.50 & \phs33 33 21.0 & $-7.9$ & 28.2 & 224 & \phm{(4),}(6)     \\
 Hercules          & 16 31 02.00 & \phs12 47 30.0 & $-6.0$ & 28.6 & 138 & \phm{(4),}(2)   
\enddata
\tablenotetext{a}{Central surface brightnesses, calculated from the
  Plummer profile fit parameters given in the discovery papers cited
  above.}
\tablenotetext{b}{The distances reported in the literature for these
  galaxies have generally been rounded off to the nearest multiple of
  10~kpc after converting from the distance modulus, which is the
  quantity directly constrained by the data.  The distances listed
  here have been calculated from the published distance moduli and
  rounded to the nearest kpc.}
\tablerefs{(1) \citealt{zucker06b}; (2) \citealt{belok07}; (3)
  \citealt{irwin07}; (4) \citealt{willman05a}; (5) \citealt{belok06};
  (6) \citealt{zucker06a}; (7) this work }
\label{targettable}
\end{deluxetable*}

In \S \ref{observations}, we describe our observations, target
selection, and data reduction, focusing in particular on our
techniques for obtaining very high precision velocity measurements
with DEIMOS.  We present the main results of this study, including
measured velocity dispersions, masses, mass-to-light ratios, and
metallicities in \S \ref{results}.  In \S \ref{discussion}, we discuss
the implications of our results for the Cold Dark Matter model and the
missing satellite problem.  We summarize our results and conclusions
in \S \ref{conclusions}.

\section{OBSERVATIONS AND DATA REDUCTION}
\label{observations}

\subsection{Observations}
\label{obs}

We obtained spectra of individual stars in eight dwarf galaxies with
the DEIMOS spectrograph \citep{faber03a} on the Keck II telescope on
2007 February 12-14.  During the observations, the weather was clear,
with seeing that varied between 0\farcs5 and 0\farcs9 (with a very
brief excursion to 1\farcs4).  The spectrograph was configured to
cover the wavelength range $6500-9000$\,\AA\ with the 1200 $\ell$/mm
grating, and the OG550 filter was used to block shorter wavelength
light.  The spectral dispersion of this setup is $0.33\,\mbox{\AA}$
per pixel, and the resulting spectral resolution, taking into account
our slit-width of 0\farcs7 and the anamorphic distortion factor of
0.7, is $1.37~\mbox{\AA}$ FWHM (corresponding to 12\,\kms\ per pixel
and 47\,\kms\ FWHM at the Ca~II triplet).
% Spectral FWHM = (0.7"/0.1185"/pix) * 0.33AA * 0.7 = 1.37 AA 
Exposures of Kr, Ar, Ne, and Xe arc lamps provided the wavelength
calibration, and an internal quartz lamp was used for flat-fielding.

We observed 18 DEIMOS slit masks, with total exposure times ranging
between 20 minutes and 2.5~hours.  One to four masks were placed on
each galaxy.  Each mask contained $\sim50-100$ stars, of which
$\sim30-80$\% were expected to be actual members of the target
galaxies from the SDSS photometry.  The positions, exposure times and
number of slits on each mask are listed in Table \ref{masktable}.
Typical target stars had magnitudes of $r \approx 20-21$.  At $r=20$,
a 1~hr exposure in good seeing conditions yields a signal-to-noise
ratio (S/N) of $\sim15$, and a 2.5~hr exposure gives a S/N of
$\sim22$, where the S/N is calculated as the average S/N per pixel in
the \ion{Ca}{2} triplet region.

\begin{deluxetable*}{lccccccc}
%\tabletypesize{\scriptsize}
\tablecaption{Keck/DEIMOS Slitmask Observing Parameters}
\tablewidth{0pt}
\tablehead{
\colhead{Mask} &
\colhead{$\alpha$ (J2000)} &
\colhead{$\delta$ (J2000)} &
\colhead{PA} &
\colhead{$t_{\rm exp}$} &
\colhead{\# of slits} &
\colhead{\% useful} \\
\colhead{name}&
\colhead{(h$\,$ $\,$m$\,$ $\,$s)} &
\colhead{($^\circ\,$ $\,'\,$ $\,''$)} &
\colhead{(deg)} &
\colhead{(sec)} &
\colhead{}&
\colhead{spectra}
}
\startdata
UMaII-1     & 08 50 38.68 & \phs63 06 45.0  & \phs\phn95.8 & 3600 &  81 & 48\%\\
UMaII-2     & 08 49 42.19 & \phs63 11 05.6  & \phs180.0 & 3600 &  87 & 52\%\\
UMaII-3     & 08 53 08.75 & \phs63 04 45.4  & \phs109.0 & 2400 &  76 & 62\%\\
UMaI-1      & 10 34 50.57 & \phs51 54 47.7  & \phs\phn65.0 & 5400 &  68 & 59\%\\
UMaI-2      & 10 34 22.23 & \phs51 56 23.9  & \phs\phn66.0 & 3600 &  62 & 65\%\\
UMaI-3      & 10 35 35.62 & \phs51 56 06.4  & \phs\phn23.3 & 5400 &  68 & 85\%\\
LeoT-1      & 09 35 00.18 & \phs17 00 56.3  & \phs\phn\phn1.0 & 3600 & 87 & 75\%\\
LeoIV-1     & 11 32 58.69 & $-$00 31 41.1   & \phs\phn\phn9.8 & 3000 & 77 & 83\%\\
ComBer-1    & 12 27 08.32 & \phs23 52 52.0  & \phs117.0 & 9000 &  78 & 62\%\\
ComBer-2    & 12 26 44.48 & \phs23 57 58.7  & \phs140.0 & 9000 &  78 & 51\%\\
ComBer-3    & 12 26 47.98 & \phs23 54 42.8  & \phn$-$20.0 & 9000 &  80 & 65\%\\ 
CVnII-1     & 12 57 12.78 & \phs34 20 43.8  & \phn$-$20.0 & 9000 &  67 & 81\%\\
CVnII-2     & 12 57 16.03 & \phs34 18 51.8  & \phs\phn50.0 & 1200 &  66 & 30\%\\
CVnI-1      & 13 27 59.38 & \phs33 34 26.8  & \phs\phn73.0  & 4140 &  91 & 87\%\\
CVnI-2      & 13 28 09.19 & \phs33 31 16.0  & \phs\phn70.5 & 4140 &  94 & 83\%\\
CVnI-3      & 13 28 14.34 & \phs33 33 23.3  & \phn\phn$-$2.0  & 4860 &  90 & 83\%\\
CVnI-4      & 13 28 02.17 & \phs33 33 36.7  & $-$112.0 & 9000 & 115 & 79\%\\
Herc-1      & 16 31 02.70 & \phs12 47 21.3  & \phs104.0 & 4500 & 106 & \phn83\%
\enddata
\tablecomments{Mask name, right ascension, declination, position angle
  and total exposure time for each Keck/DEIMOS slitmask.  The final
  two columns refer to the total number of slitlets on each mask and
  the percentage of those slitlets for which a redshift was measured.}
\label{masktable}
\end{deluxetable*} 

Target selection was carried out on star catalogs extracted from the
NYU-VAGC analysis \citep{vagc} of the Sloan Digital Sky Survey Data
Release 5 data set \citep{dr5}.\footnote{The position of Leo~T, which
  was discovered during our observing preparations, had not yet been
  processed for the VAGC at that time, so to select targets for that
  galaxy we used the standard DR5 data.}  We set the target priorities
so as to preferentially observe stars with a high likelihood of being
members of the various dwarfs based on their color, apparent
magnitude, and position.  We constructed $r,g-i$ color-magnitude
diagrams (CMDs) for each dwarf and overlaid globular cluster
isochrones from \citet[][hereafter C05]{clem05}, adjusted for the
distance reported in the literature.  We chose the best-fitting
globular cluster red giant branch (RGB) of the 3 examples provided by
C05.  We also added a horizontal branch track derived from the M13
observations of C05 and asymptotic giant branch (AGB) isochrones (for
an age of 11.2~Gyr and a metallicity of [Fe/H]$ = -1.3$ or Fe/H]$ =
  -1.7$) from \citet{girardi04}.  The highest priority targets were
  those located within 0.1 mag (in the least squares sense\footnote{As
   in \S \ref{memberselection}, when we refer to the distance between
    a star and a fiducial track in a color-magnitude diagram, we mean
    the following: $d_{\rm CMD} = \sqrt{[(g - i)_{*} - (g - i)_{\rm
          fiducial}]^{2} + (r_{*} - r_{\rm fiducial})^{2}}$, where the
    appropriate reference point for each star along the fiducial track
    is chosen so as to minimize $d_{\rm CMD}$.}) of the RGB or AGB
  tracks, or within 0.2~mag of the horizontal branch, with additional
  preference being given to brighter stars.  Stars farther from any of
  the fiducial sequences were classified as lower priority targets.
  We then designed slitmasks so as to maximize the number of high
  priority targets while still obtaining good spatial coverage.
  Slitmasks were created using the DEIMOS {\tt dsimulator} slitmask
  design software which fills in the mask area to the extent possible
  with the highest priority input targets.  This automatic selection
  was then adjusted by hand as appropriate.  The remaining space on
  the slitmasks was filled in with lower priority targets.  The slit
  width for all masks was 0\farcs7, and the minimum slit lengths were
  $\sim5$\arcsec, depending slightly on the density of target stars.

In addition to the dwarf galaxy observations, we also obtained spectra
of a radial velocity standard star, several telluric standards, and
stars in the globular cluster NGC~1904 to serve as templates for
cross-correlation with the dSph stars.  More template observations
(the globular cluster NGC~2419 and other radial velocity standards)
were obtained during additional recent Keck/DEIMOS observing runs,
with identical observing set-ups (except for the slit width).

\subsection{Data Reduction}

The data were reduced using version 1.1.4 of the DEIMOS data reduction
pipeline developed for the DEEP2 Galaxy Redshift Survey (Cooper et
al., in preparation).  Since this software was designed for faint,
resolved galaxies, we modified the pipeline to optimize reductions for
our relatively bright unresolved stellar targets.  The main
modifications were to change the cosmic ray rejection algorithm and to
allow alignment of individual 2D exposures in the spatial direction
before co-adding.  In addition, we modified the long-slit pipeline to
allow proper reduction of very bright standard stars.

\subsection{Measurement of Radial Velocities}
\label{measure_v}

We measure radial velocities by cross-correlating the observed science
spectra with a set of high signal-to-noise stellar templates.  The
stellar templates were observed with Keck/DEIMOS using the same set-up
described above.  Because template mismatch can result in significant
velocity errors, we include a wide variety of stellar types and
metallicities in our template library: giants of spectral type F8III
through M8III, subgiants and dwarf stars.  In order to cover the range
of metallicity expected in our low-luminosity dwarf galaxies, we also
include several RGB and horizontal branch (HB) stars taken from
observations of Galactic globular clusters.  The stellar templates
cover the metallicity range [Fe/H] = $-2.12$ to $+0.11$.  All science
and template spectra are rebinned onto a common wavelength array with
logarithmic wavelength bins of size 15~\kms~pixel$^{-1}$, which is
chosen to match the lowest spectral resolution present in the observed
data.

We calculate and apply a telluric correction to each velocity
measurement to account for velocity errors that result from
mis-centering an unresolved star within the slit.  Following
\citet{sohn06a}, we cross-correlate each science spectrum with a
telluric template in the regions of the strong telluric absorption:
6860-6925\,\mbox{\AA}, 7167-7320\,\mbox{\AA}, 7593-7690\,\mbox{\AA}
and 8110-8320\,\mbox{\AA}.  The telluric template was created from the
spectrum of a hot, rapidly rotating star (HR~1641, B3V) that was
allowed to drift perpendicularly across the slit (i.e., across the
0\farcs7 dimension) during the exposure, simulating a source that
uniformly fills the slit, and thus accurately reflects the mean
integrated slit function.  The mean telluric offset per mask ranged
between $-7$ and $+2$\,\kms, with a standard deviation within a mask
of 3\,\kms.  This correction is the velocity error caused by the
mis-centering of the science star within the slit from
e.g.,~astrometry errors, small mask rotation, etc.  Repeat
observations of a number of stars on multiple masks demonstrate that
the telluric correction reduces the mean absolute deviation between
independent pairs of measurements from 4.6 to 3.8~\kms, reduces the
weighted standard deviation of the velocity differences between pairs
of measurements from 5.6 to 4.2~\kms, and improves the weighted mean
difference from $-2.0$ to $-0.4$~\kms, indicating that the telluric
correction is removing both random and systematic errors from the
data.

We first calculate the telluric offset ($v_{\rm tell}$) and then
determine radial velocities ($v_{\rm obs}$) for each science spectrum.
In both cases, the template and science spectra are
continuum-subtracted; the template is then shifted and scaled to find
the best fit in reduced-$\chi^2$ space.  The final radial velocity
($v$) is then: $v = v_{\rm obs} - v_{\rm tell} - v_{\rm hel}$, where
$v_{\rm hel}$ is the heliocentric correction determined for each mask.
All the radial velocities presented in this paper include a telluric
and heliocentric correction.

The internal velocity dispersions of low-luminosity dwarf galaxies
are, in many cases, of the same order as the DEIMOS velocity errors
associated with individual measurements.  {\it In this regime, it is
  crucial to measure not only accurate velocities, but accurate
  velocity uncertainties.}  Underestimating (overestimating) the
velocity uncertainties translates directly into larger (smaller)
values of the inferred velocity dispersion using the methods described
in \S \ref{sigma}.  We determine our velocity error bars using a
Monte Carlo bootstrap method, determine the contribution from
systematic errors via repeat measurements of individual stars, and
check the precision of these errors by comparing to higher spectral
resolution data.

\begin{figure*}[t!]
\epsscale{1.0}
\plotone{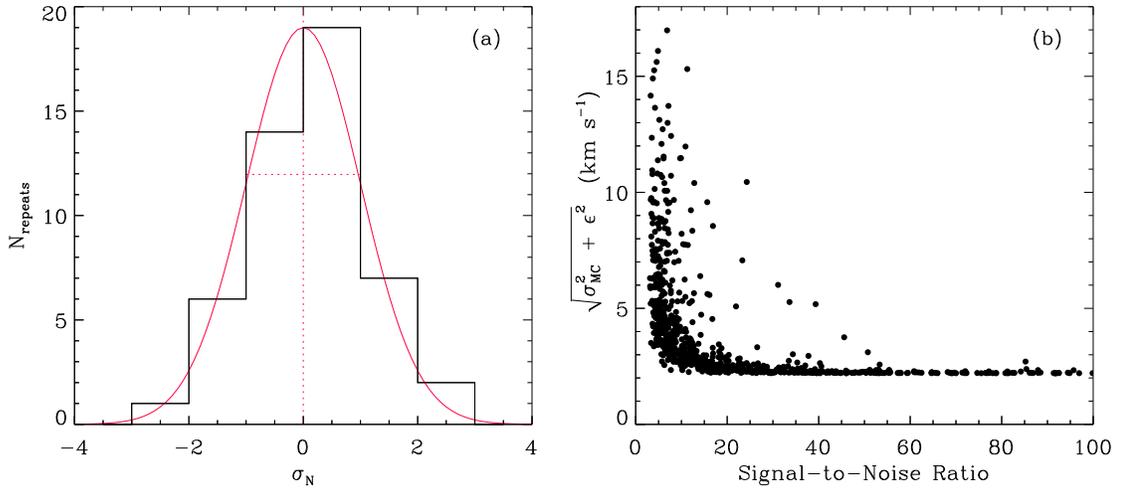}
\caption{ ({\it a}) Distribution of the normalized velocity error,
  $\sigma_N$ (as defined in Eq.~\ref{sigma_n}), for 49 pairs of
  repeated independent velocity measurements.  The best fitting
  systematic error, $\epsilon = 2.2$~\kms, is used to produce the unit
  Gaussian distribution shown in red. ({\it b}) Combined random
($\sigma_{\rm MC}$) and systematic velocity error for individual
measurements plotted as a function of the mean per pixel
signal-to-noise ratio.  These data include all of our science targets,
but not the globular cluster and standard stars observed as
spectroscopic templates.  Points that fall far from the main locus are
typically hot horizontal branch stars that lack the sharp spectral
features present in the majority of giant and dwarf stars that
comprise our sample.  Note that five stars in the sample have velocity
uncertainties larger than 18~\kms, and 62 stars have S/N greater than
100, and are therefore not displayed in this plot; we chose the axis
ranges so as to make the detailed distribution of uncertainties more
visible.}
\label{error_fig}
\end{figure*}

For the Monte Carlo method, noise is added to each pixel in the
one-dimensional spectrum of each science observation based on the
observed variance in that pixel.  We assume the variance in each pixel
is independent and distributed according to Poisson statistics.  We
then re-calculate the velocity and telluric correction for this new
spectrum using the same routines above.  Error bars are defined as the
square root of the variance in the recovered mean velocity over 500
runs of the simulations.  We next compare these Monte Carlo error
estimates to the velocity differences between independent repeat
measurements of individual stars.  Since many of our DEIMOS masks
covered overlapping sky areas, we placed 43 stars on two or more masks
to obtain multiple independent velocity measurements.  We note that
one of these stars is likely an RR Lyrae variable (\S\,\ref{uma2}) and
remove it from the sample of repeated observations.  We are left with
49 pairs of independent velocity measurements.  The velocity
difference between these independent observations samples the `true'
error distribution.  We define a normalized error ($\sigma_N$) as the
velocity difference between two independent measurements ($v_1$,
$v_2$), divided by the quadrature sum of all error contributions:
\begin{equation}
\sigma_{\rm N} = \frac{v_1 - v_2}{\sqrt{\sigma_{\rm MC1}^2 +
    \sigma_{\rm MC2}^2 + 2\epsilon^2}},
\label{sigma_n}
\end{equation}
\noindent
where the Monte Carlo errors ($\sigma_{\rm MC1}$, $\sigma_{\rm MC2}$)
on each measurement are combined in quadrature with an additional term
$\epsilon$ equal to the error contribution from systematics not
accounted for in the Monte Carlo simulation.  The $\sigma_{\rm N}$
distribution should be a Gaussian of unit width.  We therefore
determine the unknown contribution from other systematic errors by
fitting the parameter $\epsilon$ to produce a unit Gaussian
distribution.  In the left panel of Figure~\ref{error_fig}, the final
$\sigma_{\rm N}$ distribution is plotted for the best fitting value of
$\epsilon = 2.2$\,\kms.  The final velocity errors used in our
analysis are the quadrature sum of the Monte Carlo and systematic
errors.  In the right panel of Figure~\ref{error_fig}, we plot the
final velocity errors as a function of the mean per pixel
signal-to-noise ratio for all the individual stellar velocities
presented in this paper.  Stars that fall far from the main locus of
points are typically hot horizontal branch stars (which have few sharp
spectral features).  The median velocity uncertainty for our sample of
member stars in the ultra-faint dwarfs is 3.4~\kms; including mask
alignment stars and bright foreground stars that tend to have higher
S/N, the median uncertainty for the entire data set is 2.7~\kms.

To demonstrate our ability to accurately measure velocities and
recover velocity dispersions, we compare our observations of the
Galactic globular cluster NGC~2419 to higher spectral resolution
Keck/HIRES observations of the same cluster (P.~C{\^ o}t{\' e} 2007,
private communication).  We measure radial velocities for 26~stars
between $1\arcmin$ and $4\arcmin$ of the cluster center.  The HIRES
data contain a similar number of stars in this region, although very
few stars overlap between the two data sets.  The HIRES spectrograph
has a high spectral dispersion (0.02\,\AA\ per pixel) and much more
accurate individual velocity measurements ($\sim0.95$~\kms).  We
compute the recession velocity and velocity dispersion of NGC~2419 for
both data sets using the maximum-likelihood technique described in \S
\ref{sigma}.  For the recession velocity, we measure $\langle\hat
u\rangle_{\rm DEIMOS} = -20.7 \pm 0.6$~\kms\ compared to $\langle\hat
u\rangle_{\rm HIRES} = -21.2 \pm 0.5$~\kms, and for the velocity
dispersion, $\sigma_{\rm DEIMOS} = 2.3 \pm 0.4$~\kms\ compared to
$\sigma_{\rm HIRES} =2.3 \pm 0.3$\,\kms.  The DEIMOS observations
agree within the 1-$\sigma$ limits of the more accurate HIRES
measurements.  Both sets of measurements agree with the published
values for this cluster \citep{pryor93}.  While our NGC~2419
observations have somewhat higher signal-to-noise ratios than those
typical of our dwarf galaxy observations, this comparison demonstrates
that we are able to reliably measure the kinematics in systems with
extremely low velocity dispersions (smaller than expected for the
dwarf galaxies).  We also note that we measure a metallicity based on
the Ca triplet lines (\S \ref{measure_EW}) for NGC~2419 of [Fe/H]$ =
-2.0$, and the data are consistent with no intrinsic metallicity
spread within the cluster.  The standard metallicity for this cluster
is [Fe/H]$ = -2.12$ \citep{harris96}.

As a further test of our ability to measure reliable velocities, we
compare our observations to high resolution spectroscopy in UMa~I by
\citet{kleyna05}.  These authors presented Keck/HIRES spectra of 7
stars in the UMa~I region (5 members and 2 non-members).  We
re-observed all 7 stars with multiple measurements for 2
stars.\footnote{Note that the SDSS DR2 coordinates given in
  \citet{kleyna05} for the target stars are up to 8\arcsec\ off from
  the true positions as given in the DR5 data or on Palomar Sky Survey
  plates.}  We find excellent agreement between our measurements and
those of \citeauthor{kleyna05}~for 6 of the stars (differences within
the 1~$\sigma$ uncertainties for 5 out of 6 and less than
1.7~\kms\ for all 6); for star 7, both of our measurements are
significantly discrepant with \citeauthor{kleyna05}'s velocity.  Star
7 had the lowest S/N in \citeauthor{kleyna05}'s observations, and they
described the Ca triplet lines as ``barely discernible above the
noise.''  We conclude that either this star is a binary or variable
star, or the velocity measured by \citet{kleyna05} is in error by
$\sim8$~\kms\ (their estimated uncertainty is 5~\kms).

Radial velocities were successfully measured for 1015 of the 1460
extracted spectra across the 18~observed science masks.  This total
includes 50 duplicate measurements of individual stars and 124
objects identified as galaxies or quasars.  The latter objects will be
very useful as background objects for proper motion studies and will
be the subject of a future paper.  The majority of spectra for which
we could not measure a redshift did not have sufficient S/N.  The
fitted velocities are visually inspected to insure the reliability of
the measured redshift and the overall quality of the spectrum.  The
final sample of stellar radial velocities consists of 841~unique
measurements across the eight target dwarf galaxies.

\subsection{Measurement of Equivalent Widths and Metallicities}
\label{measure_EW}

We estimate the metallicity ([Fe/H]) of individual RGB stars in our
target galaxies using the \ion{Ca}{2} triplet absorption lines near
$\lambda = 8500$\,\mbox{\AA}. We calculate the equivalent widths (EWs)
of the three \ion{Ca}{2} absorption lines using the line and continuum
definitions of \citet{rutledge97a}.  The three EWs are combined into a
single quantity as $\Sigma {\rm Ca} = 0.5{\rm
  EW}(\lambda8498\mbox{\AA}) + 1.0{\rm EW}(\lambda 8542\mbox{\AA}) +
0.6{\rm EW}(\lambda8662\mbox{\AA})$.  We determine the error on this
combined quantity with the Monte Carlo method described above.  Added
in quadrature to the Monte Carlo uncertainties is a systematic
uncertainty of $0.3$\,\mbox{\AA}, which we determined from repeat
measurements as described in \S \ref{measure_v}.  We convert
$\Sigma{\rm Ca}$ into metallicity using the \citet*{rutledge97b}
empirical calibration relationship:
\begin{equation}
[\mbox{Fe/H}] = -2.66 + 0.42 [\Sigma {\rm Ca} - 0.64 (V_{\rm HB} - V)]
\end{equation}
\noindent 
The term $(V_{\rm HB} - V)$ is the magnitude difference between the
horizontal branch and the observed star, and corrects for surface
gravity effects.  We assume an absolute magnitude for a metal-poor
horizontal branch $M_{V,\rm HB} = 0.88$ \citep{clem05}, and calculate
the apparent magnitude, $V_{\rm HB}$, using the distance modulus of
each galaxy (see Table \ref{targettable}).  The uncertainties in the
distance moduli are included in the total metallicity uncertainties we
derive.  Note that assuming a single value for the horizontal branch
magnitude in the possible presence of multiple stellar populations may
add an additional $\sim 0.07$~dex to the metallicity uncertainties
\citep{koch06}.  We convert the SDSS $g$-band magnitudes into $V$-band
using the photometric transformations of \citet{blanton07} and
reddening corrections from \citet*{sfd98}.  The
\citeauthor{rutledge97b}~calibration relation is derived for RGB stars
in Milky Way globular clusters using the abundance scale of
\citet{cg97}, and while the calibration data only extend to [Fe/H]$ =
-2.1$ it is reasonable to extrapolate the relation to the slightly
lower metallicities found in some of our low-luminosity dwarfs (see \S
\ref{metals}).  We restrict our metallicity analysis in \S
\ref{metals} to only the RGB stars in the dwarf galaxies.

To remove foreground dwarf stars from the sample, we will use the
equivalent width of the \ion{Na}{1} $\lambda\lambda
8183,8195$~\AA\ absorption lines, which are strongly dependent on
surface gravity and temperature \citep{spinrad71,schiavon97}.  We
measure the \ion{Na}{1} equivalent width using the line and continuum
definitions of \citet{schiavon97}.  \citeauthor{schiavon97}~show that
the \ion{Na}{1} EW is expected to be 1~\AA\ or greater in M-type dwarf
stars, whereas this feature is much weaker in giant stars at the same
temperature. \citet{gilbert06} have used this feature to successfully
discriminate between dwarf and giant stars for a similar spectroscopic
sample.

\section{RESULTS}
\label{results}

\subsection{Selection of Members}
\label{memberselection}

We use two complementary techniques to determine which of the observed
stars are members of the dwarf galaxies and which are foreground
stars.  The first method classifies stars based on objective criteria:
velocity, distance from the fiducial RGB and HB tracks (corrected for
foreground extinction using the reddenings from \citealt{sfd98} and
for the distance of the galaxy), and the equivalent width of the
\ion{Na}{1} 8190~\AA\ absorption lines (described above).  The exact
cutoffs for each of these parameters needed to be adjusted in a few
cases, but in general we use a 3~$\sigma$ cutoff in velocity
(requiring a prior iteration to estimate the velocity dispersion),
color-magnitude distance limits (defined as in \S \ref{obs}) of
0.2~mag for RGB stars and 0.4~mag for HB stars, and a \ion{Na}{1}
equivalent width of less than 1.0~\AA.  Notable exceptions to these
cutoffs include Coma Berenices, which is located so nearby that we
detect a number of subgiants, blue stragglers, and main sequence stars
at $r > 21.5$ --- at these faint magnitudes we extend the allowed
distance from the fiducial CMD track to 0.5~mag; and CVn~I, which has
a broad giant branch that also necessitates widening the cutoff
distance from the RGB track.  In addition, CVn~I has so many member
stars (214) that the presence of a 3~$\sigma$ outlier is likely (and
indeed we find one), so the velocity cutoff must be extended to
3.5~$\sigma$, where there is only a 10\% chance of finding a member
star in our sample.

\begin{figure*}[t!]
\epsscale{1.20}
\plotone{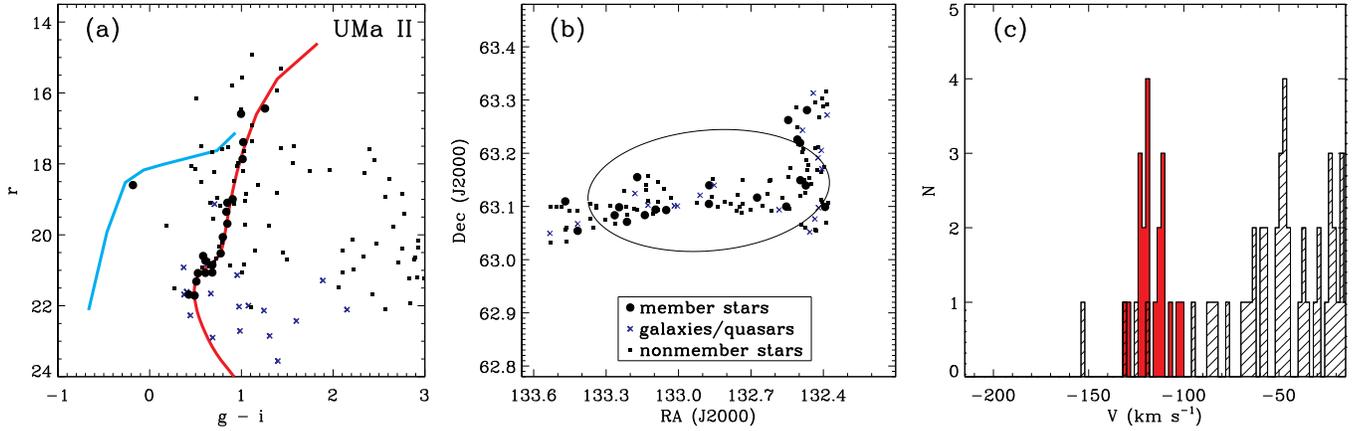}
\caption{(\emph{a}) Color-magnitude diagram of observed stars in Ursa
  Major~II.  The large black circles represent stars identified as
  radial velocity members of the galaxy, the small black dots
  represent stars identified as non-members, and the blue crosses are
  spectroscopically confirmed background galaxies and quasars.  The
  red curve shows the location of the red giant branch, subgiant
  branch, and main sequence turnoff populations in the globular
  cluster M92 and the blue curve shows the location of the horizontal
  branch of M13, both corrected for Galactic extinction and shifted to
  a distance of 32 kpc \citep[data from][]{clem05}.  (\emph{b})
  Spatial distribution of observed stars in Ursa Major~II.  Symbols
  are the same as in (\emph{a}) (the figure legend applies to both
  panels), and the ellipse represents the half-light radius of UMa~II
  from \citet{zucker06b}.  (\emph{c}) Velocity histogram of observed
  stars in Ursa Major~II.  Velocities are corrected to the
  heliocentric rest frame.  The filled red histogram represents stars
  classified as members, and the hatched black-and-white histogram
  represents non-members.  The velocity bins are 2~\kms\ wide.  }
\label{uma2_obsplot}
\end{figure*}

\begin{figure*}[t!]
\epsscale{1.20}
\plotone{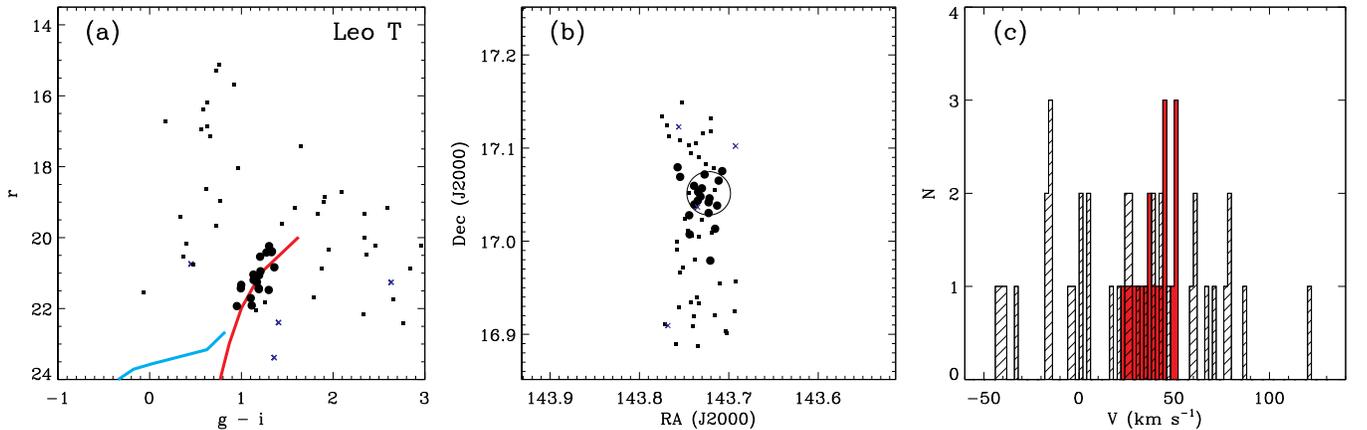}
\caption{Same as Figure \ref{uma2_obsplot}, but for Leo~T.}
\label{leot_obsplot}
\end{figure*}

\begin{figure*}[t!]
\epsscale{1.20}
\plotone{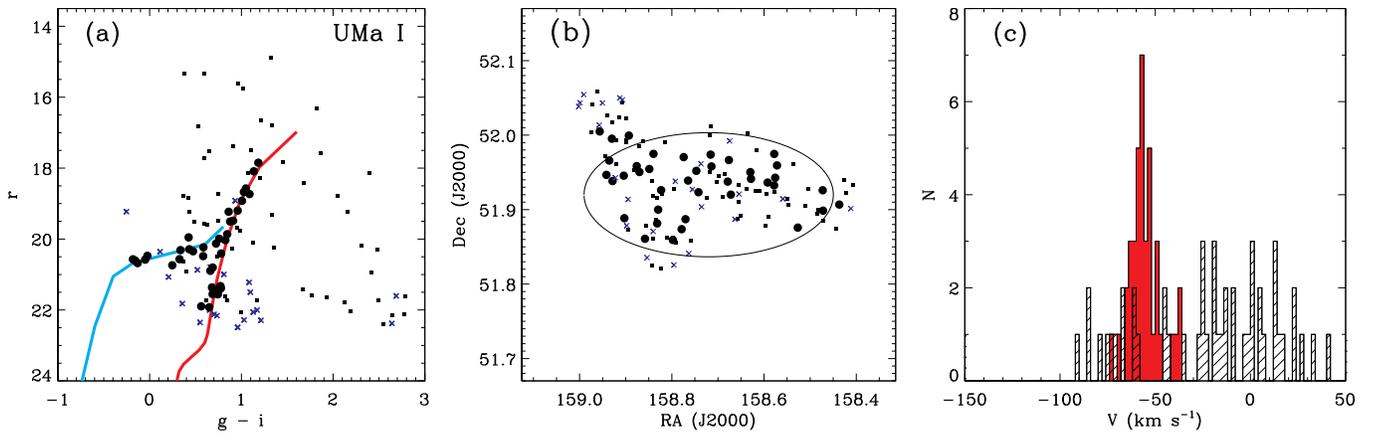}
\caption{Same as Figure \ref{uma2_obsplot}, but for Ursa Major~I.}
\label{uma1_obsplot}
\end{figure*}

\begin{figure*}[t!]
\epsscale{1.20}
\plotone{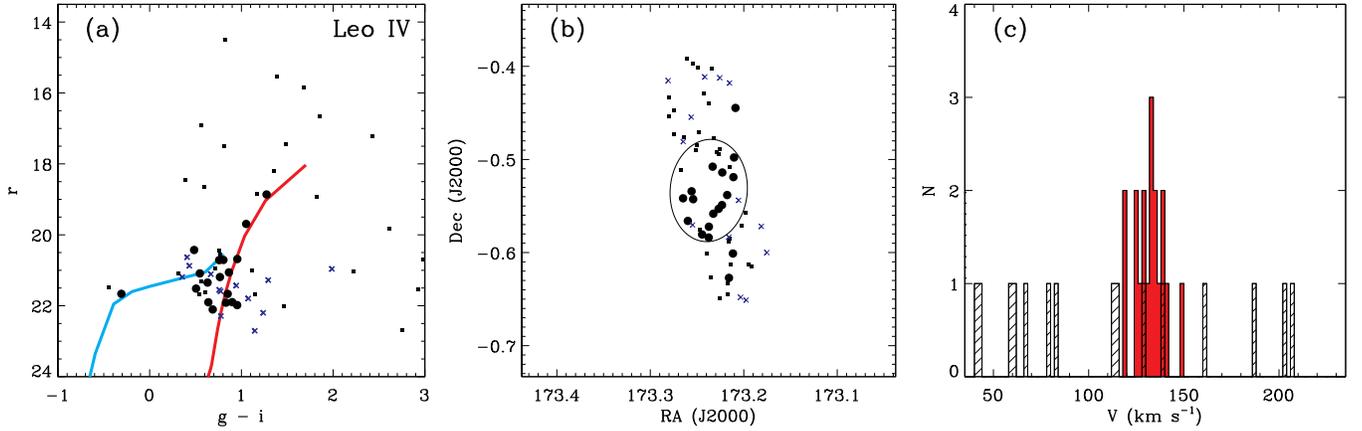}
\caption{Same as Figure \ref{uma2_obsplot}, but for Leo~IV.}
\label{leo4_obsplot}
\end{figure*}

\begin{figure*}[t!]
\epsscale{1.20}
\plotone{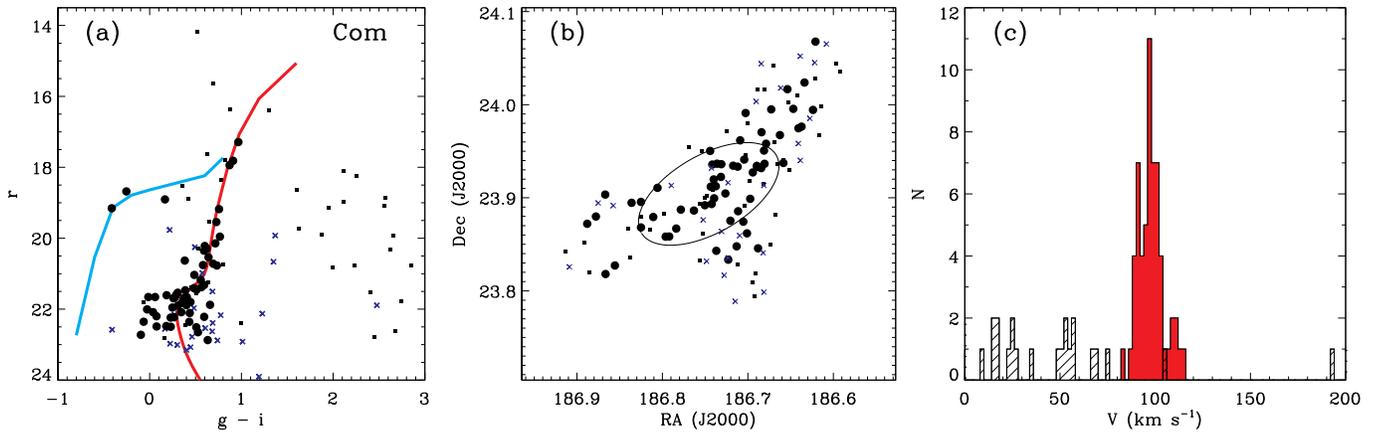}
\caption{Same as Figure \ref{uma2_obsplot}, but for Coma Berenices.}
\label{coma_obsplot}
\end{figure*}

\begin{figure*}[t!]
\epsscale{1.20}
\plotone{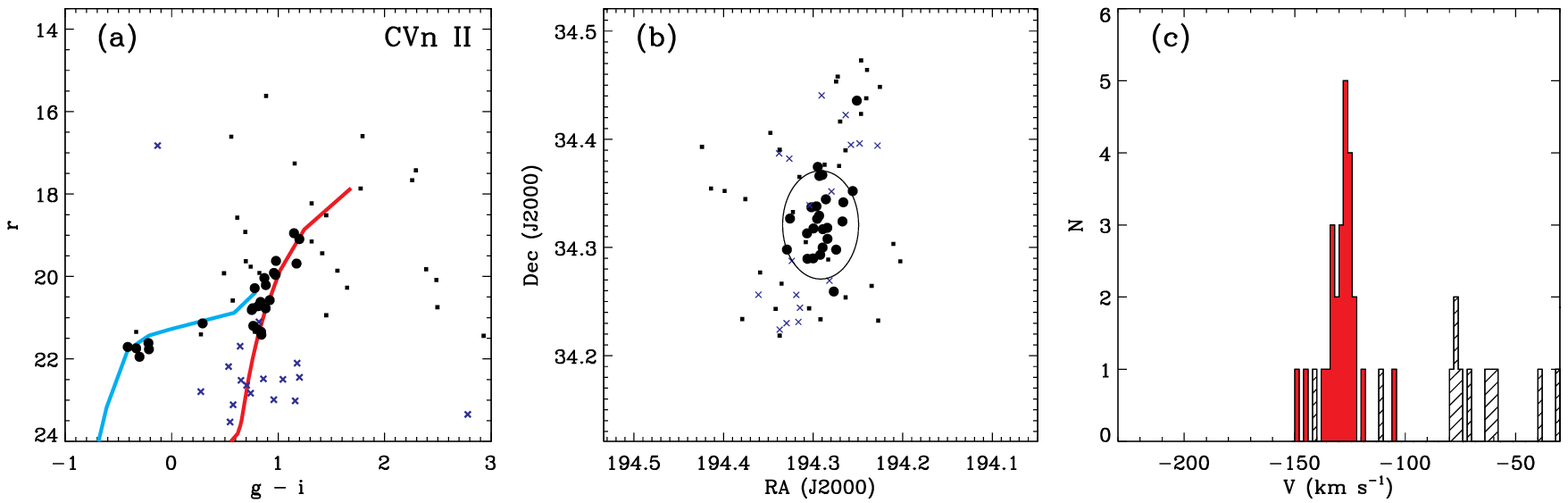}
\caption{Same as Figure \ref{uma2_obsplot}, but for Canes Venatici~II.}
\label{cvn2_obsplot}
\end{figure*}

\begin{figure*}[t!]
\epsscale{1.20}
\plotone{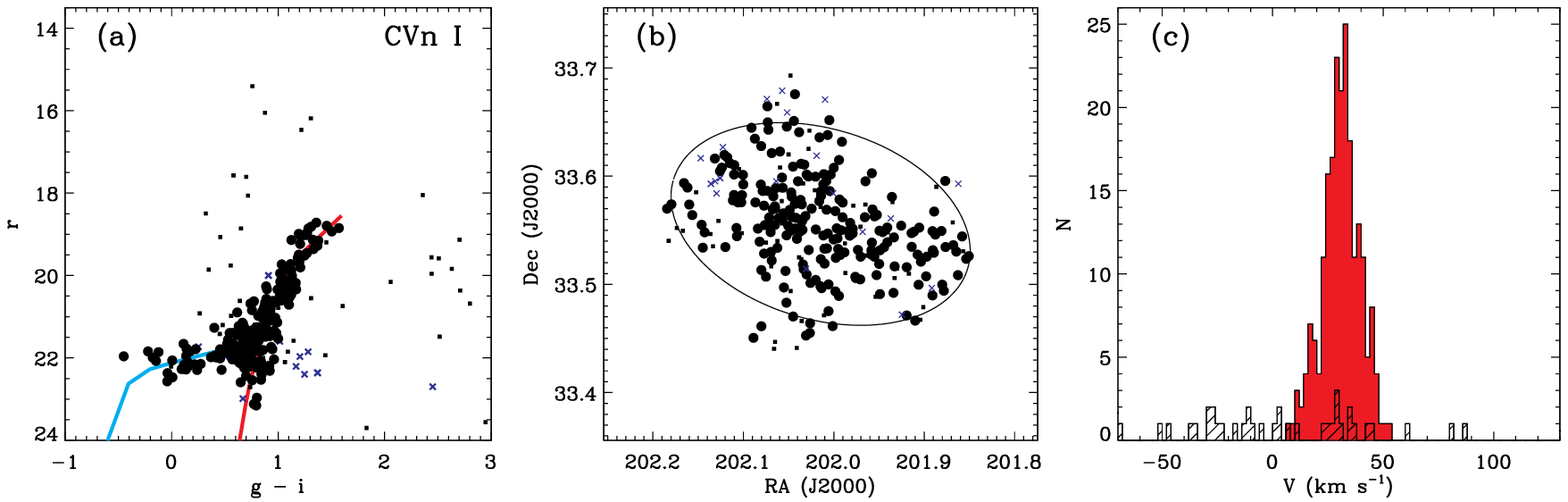}
\caption{Same as Figure \ref{uma2_obsplot}, but for Canes Venatici~I.}
\label{cvn1_obsplot}
\end{figure*}

\begin{figure*}[t!]
\epsscale{1.20}
\plotone{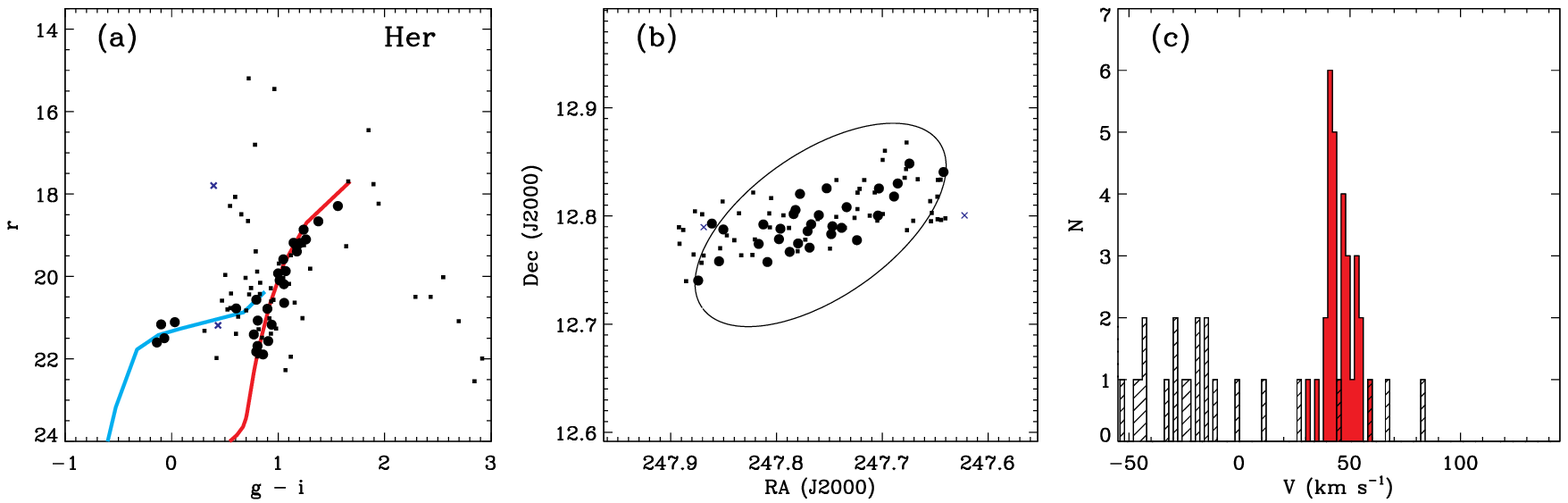}
\caption{Same as Figure \ref{uma2_obsplot}, but for Hercules.}
\label{herc_obsplot}
\end{figure*}

The second method is to examine individually the following properties
of each star: velocity, location in the CMD, spatial position, fitted
spectral type, metallicity, \ion{Na}{1} equivalent width, and if
necessary, the spectrum.  Combining all of the available information
about each star, and using thresholds similar to those described above
(but less rigid), we classify each star as a likely member or
non-member ``by eye''.  Both of these methods are similar in spirit to
the techniques described by \citet{gilbert06} and \citet{raja06} to
separate M31 red giants from foreground main sequence stars, but
without employing a full maximum-likelihood calculation, which is not
necessary for these data because the dSph stars are more localized in
parameter space than M31 halo stars are, and because of the higher
S/N.  In all cases, we find excellent agreement between the member
samples identified with the two methods (with occasional threshold
tweaks required to produce a perfect match).  A few of the galaxies
contain questionable member stars that significantly affect the
derived velocity dispersions, and these cases will be discussed
individually in \S \ref{comments}.  We display color-magnitude
diagrams and spatial distributions for the observed stars in each
dwarf galaxy in Figures \ref{uma2_obsplot} - \ref{herc_obsplot}.

\subsection{Central Velocity Dispersions}
\label{sigma}

Given the member samples selected in the previous subsection, we use
the maximum-likelihood method described by \citet{walker06} to
calculate simultaneously the mean velocities and velocity dispersions
of each galaxy.\footnote{Note that the numerical values of the
  parameters $a$ and $b$ in Equation 9 of \citeauthor{walker06}~are
  negative, since they are proportional to the second derivatives of
  $\ln{p}$ evaluated at the maximum of the function.  The
  uncertainties on the mean velocity and the velocity dispersion
  should therefore be defined as $d\langle\hat u\rangle = \sqrt{\left|
    a \right|}$ and $d\sigma = \sqrt{\left|b\right|}$ to avoid
  imaginary results.}  This method assumes that the observed velocity
dispersion is the sum of the intrinsic galaxy dispersion and the
dispersion produced by measurement errors, as well as that the
velocity distribution is reasonably approximated by a Gaussian.  The
derived velocities and intrinsic velocity dispersions are displayed in
Table \ref{dispersiontable}.  We find dispersions ranging from
$3.3\pm1.7$~\kms\ for Leo~IV to $7.6\pm0.4$~\kms\ for Canes
Venatici~I.

\begin{deluxetable*}{l c c c c c c}
\tablewidth{0pt}
\tablecolumns{7}
\tablecaption{Radial Velocities and Velocity Dispersions}
\tablehead{
\colhead{Galaxy} & \colhead{$\langle\hat u\rangle_{hel}$}  & 
\colhead{$d\langle\hat u\rangle_{hel}$} & 
\colhead{$\langle\hat u\rangle_{GSR}$} & 
\colhead{$\sigma$} & \colhead{$d\sigma$} & \colhead{Number}  \\
\colhead{} & \colhead{(\kms)} & \colhead{(\kms)} &
\colhead{(\kms)} & \colhead{(\kms)} & \colhead{(\kms)} & \colhead{of stars}
}
\startdata 
Ursa Major II     & $-$116.5      & 1.9 & \phn$-$33.4    & 6.7 & 1.4 & \phn20 \\
Leo T             & \phn\phs38.1  & 2.0 & \phn$-$58.4    & 7.5 & 1.6 & \phn19 \\
Ursa Major I      & \phn$-$55.3   & 1.4 & \phn\phn$-$7.1 & 7.6 & 1.0 & \phn39 \\
Leo IV            & \phs132.3     & 1.4 & \phn\phs10.1   & 3.3 & 1.7 & \phn18 \\
Coma Berenices    & \phn\phs98.1  & 0.9 & \phn\phs81.7   & 4.6 & 0.8 & \phn59 \\
Canes Venatici II & $-$128.9      & 1.2 & \phn$-$95.5    & 4.6 & 1.0 & \phn25 \\
Canes Venatici I  & \phn\phs30.9  & 0.6 & \phn\phs77.6   & 7.6 & 0.4 &    214 \\
Hercules          & \phn\phs45.0  & 1.1 & \phs144.6      & 5.1 & 0.9 & \phn30    
\enddata
\label{dispersiontable}
\end{deluxetable*}

\begin{figure*}[t]
\epsscale{1.1}
\plottwo{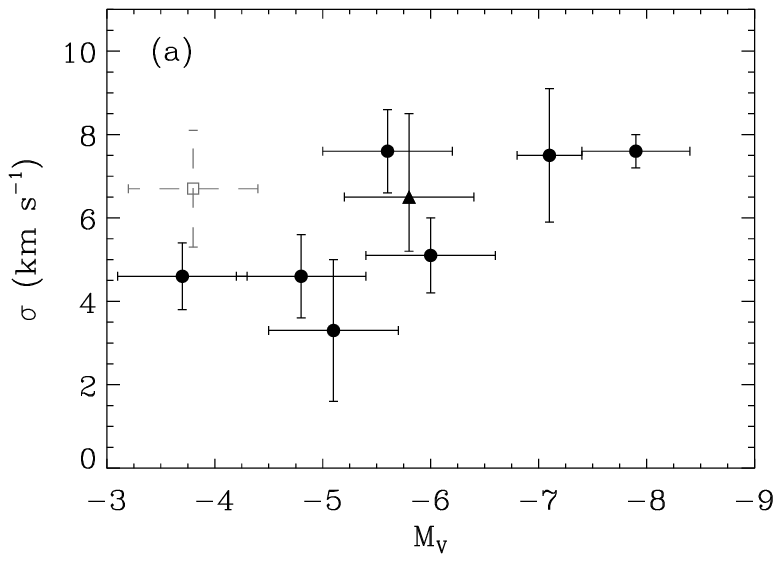}{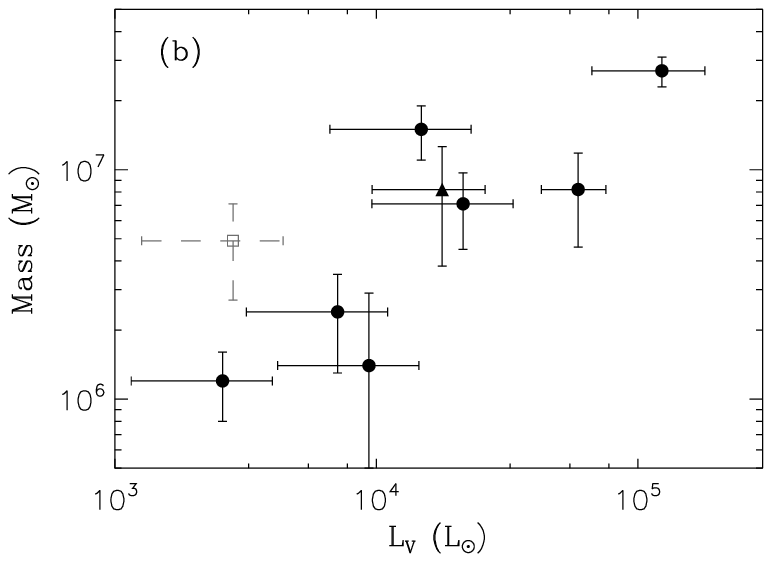}
\caption{(\emph{a}) Velocity dispersion as a function of absolute
  magnitude for the ultra-faint dwarfs.  The filled black symbols
  represent the gravitationally bound dwarfs and the open gray symbol
  represents UMa~II, which is thought to be tidally disrupted (see \S
  \ref{tidal}).  Circles are ultra-faint dwarfs in this sample and the
  triangle is the Bo{\"o}tes dSph \citep{martin07}.  (\emph{b})
  Dynamical mass as a function of total $V$-band luminosity.  Symbols
  are the same as panel (\emph{a}).  The ultra-faint dwarf galaxies
  clearly display a trend in which the more luminous galaxies have
  larger velocity dispersions and correspondingly larger masses.
  Perhaps surprisingly, there appears to be a simple power-law
  relationship between mass and luminosity.}
\label{sigmaplot}
\end{figure*}

We plot the stellar velocity dispersions as a function of absolute
magnitude in Figure \ref{sigmaplot}\emph{a}.  There is a significant
correlation of velocity dispersion with absolute magnitude, with the
more luminous galaxies ($M_{V} \simless -6$) having larger dispersions
of $\sim7-8$~\kms\ and the fainter galaxies ($M_{V} \simgtr -6$)
exhibiting smaller dispersions of $\sim4-5$~\kms.  The four
low-luminosity galaxies Coma Berenices, CVn~II, Hercules, and Leo~IV
are the first galaxies to break the velocity dispersion ``barrier'' at
$\sim7$~\kms\ that observations of the previously known dSphs had
suggested \citep{gilmore07}.  The unprecedentedly low velocity
dispersions of these galaxies, and the correlation with absolute
magnitude down to such low luminosities, demonstrate that if there is
a floor on the masses of dSphs, it does not appear to have been
reached yet.

The likely presence of unresolved binary stars in our stellar velocity
sample may increase the measured velocity dispersion of our target
galaxies due to binary orbital motion.  \citet{olszewski96} simulated
the effect of binaries on the velocity dispersions of the Draco and
Ursa Minor dSphs with very similar sample sizes and velocity
uncertainties as the present study.  Assuming the binary fractions
determined for Draco and Ursa Minor (which range between 0.2 and 0.3
for relevant binary periods), \citeauthor{olszewski96}~suggested that
the velocity dispersion from binaries alone is on the order of
1.5~\kms.  Since it is possible that the binary fractions may be
different in the lower luminosity galaxies we observed, we use this
estimate only as a guide.  For the highest velocity dispersion systems
listed in Table~\ref{dispersiontable}, the effect of binaries is
negligible.  This result is consistent with conclusions from previous
groups for other Local Group dSphs \citep{kleyna99, walker06}.  For
our lowest dispersion system, Leo~IV, the
\citeauthor{olszewski96}~binary correction would reduce the measured
dispersion from 3.3\,\kms\ to 2.9\,\kms.  However, this difference is
significantly smaller than our measurement uncertainty of 1.7\,\kms,
so we do not correct our measured dispersions for the presence of
binaries.  Unless the binary star fraction in these ultra-low
luminosity dwarfs is significantly larger than that of other dwarf
galaxies, binaries do not significantly inflate the measured
dispersions and inferred masses of even the lowest-dispersion dwarf
galaxies in our sample.

\subsection{Total Masses}
\label{masses}

The process of determining the total mass of a dwarf spheroidal galaxy
from the velocities of a relatively modest sample of stars that are
probably located well inside the virial radius of the galaxy's dark
matter halo is fraught with difficulty.  The standard technique in the
literature is to assume that (1) the galaxy is spherical, (2) the
galaxy is in dynamical equilibrium, (3) the galaxy has an isotropic
velocity dispersion, and (4) the light distribution of the galaxy
traces its mass distribution.  All four of these assumptions may be
false in reality, especially for the ultra-faint dwarfs that are the
subject of this paper.  SDSS photometry and followup imaging reveal
that most of the dwarfs are elongated, demonstrating that they are not
spherically symmetric systems and probably do not have isotropic
velocity dispersion tensors.  Several of the dwarfs also appear
irregular, opening up the possibility that their structure has been
significantly affected by the tidal field of the Milky Way.  However,
these apparently irregular isodensity contours could also be the
result of the extremely low surface densities of the galaxies, which
make their stellar distributions difficult to determine accurately.
Finally, the nearly flat velocity dispersion profiles observed in all
of the dSphs where spatially resolved kinematics are available
indicate that light does not trace mass \citep{walker06,wu07}.
Despite these objections, the samples of stars in the ultra-faint
dwarfs that are spectroscopically accessible with current instruments
are so small that more sophisticated analyses are not possible (with
the exception of CVn~I, which will be discussed in more detail in a
future paper).  We therefore use the method of \citet{illingworth76}
to estimate total masses for the observed galaxies:

\begin{equation}
M_{tot} = 167 \beta r_{c} \sigma^{2},
\end{equation}

\noindent
where $\beta$ is a parameter that depends on the concentration of the
system and is generally assumed to be 8 for dSphs \citep{mateo98},
$r_{c}$ is the \citet{king62} profile core radius, and $\sigma$ is the
observed central velocity dispersion.  For most of the new dwarfs,
only Plummer (half-light) radii rather than King core radii are
available in the literature, but we can use the fact that $r_{c} =
0.64*r_{\rm Plummer}$ to estimate the King radii.  The radii and
luminosities we have assumed for these calculations are given in
Appendix \ref{data}.  Our derived masses for each galaxy are listed in
Table \ref{masstable} and plotted in Figure~\ref{sigmaplot}\emph{b}.
We note that objects in the bottom left corner of the plot are both
the least massive and least luminous known galactic systems.

\begin{deluxetable*}{l c c c c}
\tablewidth{0pt}
\tablecolumns{5}
\tablecaption{Masses, Mass-to-Light Ratios, and Metallicities}
\tablehead{
\colhead{Galaxy} & \colhead{Mass}  & 
\colhead{M/L$_{V}$} & \colhead{[Fe/H]} & 
\colhead{$\sigma_{[Fe/H]}$} \\
\colhead{} & \colhead{(\msun)} & 
\colhead{(\msun/\lsun)} & \colhead{} & 
\colhead{}
}
\startdata 
Ursa Major II\tablenotemark{a} & $4.9 \pm 2.2 \times 10^{6}$ & $1722 \pm 1226$ & $-1.97 \pm 0.15$ & 0.28 \\
Leo T             & $8.2 \pm 3.6 \times 10^{6}$ & \phn$138 \pm \phn71$  & $-2.29 \pm 0.10$ & 0.35 \\
Ursa Major I      & $1.5 \pm 0.4 \times 10^{7}$ & $1024 \pm 636$  & $-2.06 \pm 0.10$ & 0.46 \\
Leo IV            & $1.4 \pm 1.5 \times 10^{6}$ & \phn$151 \pm 177$  & $-2.31 \pm 0.10$ & 0.15 \\
Coma Berenices    & $1.2 \pm 0.4 \times 10^{6}$ & \phn$448 \pm 297$  & $-2.00 \pm 0.07$ & 0.00 \\
Canes Venatici II & $2.4 \pm 1.1 \times 10^{6}$ & \phn$336 \pm 240$  & $-2.31 \pm 0.12$ & 0.47 \\
Canes Venatici I  & $2.7 \pm 0.4 \times 10^{7}$ & \phn$221 \pm 108$  & $-2.09 \pm 0.02$ & 0.23 \\
Hercules          & $7.1 \pm 2.6 \times 10^{6}$ & \phn$332 \pm 221$  & $-2.27 \pm 0.07$ & 0.31
\enddata
\label{masstable}
\tablenotetext{a}{UMa~II may be a tidally disrupted remnant, which
  would artificially inflate its mass and mass-to-light ratio.}
\end{deluxetable*}

The ultra-faint Milky Way satellites have masses ranging from just
over $10^{6}$~\msun\ (Coma Berenices) up to $2.8 \times
10^{7}$~\msun\ (Canes Venatici~I).  Not surprisingly, CVn~I, which is
nearly as bright as previously known dSphs such as Ursa Minor and
Draco, has a mass that is similar to those of the original Milky Way
dSphs.  Combining the measured masses with the absolute magnitudes
listed in Table \ref{targettable}, we can calculate $V$-band
mass-to-light ratios, which are presented in Table \ref{masstable}.
The new dwarfs continue the trend of an anti-correlation between
luminosity and M/L that has been known for many years
\citep[e.g.,][]{mateo93}, reaching mass-to-light ratios of $\sim1000$
in $V$-band solar units.  Although the uncertainties on the
mass-to-light ratios are substantial, owing primarily to the poorly
known luminosities of the ultra-faint dwarfs, it is clear that all of
these galaxies have quite large mass-to-light ratios.  The existence
of galaxies with similar properties to these was predicted recently by
\citet{rg05} and \citet*{read06}, but the measured masses seem to be in
better agreement with the models of \citet{rg05}.

\subsection{Metallicities}
\label{metals}

The mean stellar metallicity of a galaxy reflects the enrichment
history of the interstellar medium at the time the stars were formed.
We determine the mean metallicity, [Fe/H], for the new dwarf galaxies
based on the \ion{Ca}{2} triplet equivalent width (\S
\ref{measure_EW}).  While we can reliably measure equivalent widths
for the majority of our target stars, the \citet{rutledge97b}
empirical calibration that we use to convert to [Fe/H] is valid only
for RGB stars.  We therefore only include stars brighter than $M_V =
+1.5$ and redder than $(g-r) > 0.3$ (to avoid HB stars) in the
metallicity analysis.  We determine the mean metallicity and
metallicity spread using the maximum-likelihood technique described in
\S \ref{sigma}.  While the metallicity distributions are not
necessarily Gaussian, as the maximum-likelihood calculation assumes,
we find that the mean and median of the observed metallicity
distributions give similar results.  We run the maximum-likelihood
algorithm twice, rejecting 3~$\sigma$ outliers on the second run.  The
mean metallicities and metallicity spreads are listed in
Table~\ref{masstable}.  We find metallicities ranging from [Fe/H]
=$-1.97 \pm 0.15$ for UMa~II down to [Fe/H]=$-2.31$ for CVn~II and
Leo~IV.  We note that several of our galaxies have mean metallicities
equal to those of the most metal-poor globular clusters and lower than
those of other dwarf galaxies \citep{harris96,mateo98}, making them,
along with the Bo{\" o}tes dSph \citep{munoz06}, the most metal-poor
stellar systems known.

\begin{figure*}[t!]
\epsscale{1.0}
\plotone{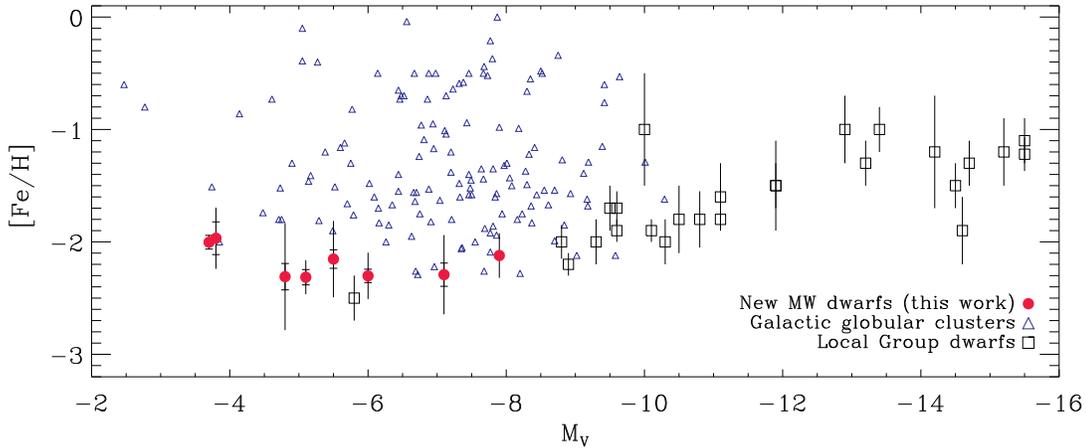}
\caption{Metallicity-luminosity relationship for dwarf galaxies in the
  Local Group.  The new ultra-faint galaxies (red circles) follow the
  trend of decreasing metallicity with luminosity set by more luminous
  dwarf galaxies (black squares).  The two lowest-luminosity objects
  (UMa~II and Com) show possible evidence of tidal stripping.  In
  comparison, Galactic globular clusters (blue triangles) do not
  follow any luminosity-metallicity relationship.  Data for luminous
  dwarf galaxies are from \citet{mateo98}, Galactic globular clusters
  from \citet{harris96} and the ultra-low luminosity dwarf Bo{\"o}tes
  (open square at $M_V = -5.8$) from \citet{munoz06}.
  \citet{martin07} find a somewhat higher metallicity for Bo{\"o}tes
  of [Fe/H]$ = -2.1$.  The smaller horizontal bars on our galaxy
  measurements represent the uncertainty in the mean metallicity;
  internal metallicity spreads are indicated by the larger vertical
  bars.  The ultra-low luminosity dwarfs are among the most metal-poor
  stellar systems in the known universe.}
\label{feh_fig}
\end{figure*}

A strong correlation exists between the total luminosity of dwarf
galaxies in the Local Group and their stellar metallicities
\citep{mateo98, grebel03}.  In comparison, Galactic globular clusters
follow no such relationship \citep{harris96}.  In
Figure~\ref{feh_fig}, we show that all but the two faintest of the
ultra-low luminosity galaxies follow the luminosity-metallicity
relationship defined by the more luminous dSphs.  The two galaxies
deviating from this relationship are UMa~II and Com.  These are the
nearest as well as the lowest-luminosity objects in our sample and
both galaxies (particularly UMa~II) show a variety of evidence
suggesting that they are undergoing tidal disruption by the Milky Way.
As discussed in \S\,\ref{tidal}, we interpret the high metallicities
in these two objects as evidence that their formation mass may have
been significantly larger than their present mass.

The ultra-low luminosity galaxies extend the luminosity-metallicity
relation in the Local Group by an additional four magnitudes to $M_V =
-4.8$.  The location of these seven objects (including Bo{\"o}tes) on
the same relationship defined by brighter dSphs is significant.  It
suggests that the stars formed in these galaxies are connected to the
{\it present} mass of the galaxy and argues against significant tidal
stripping, unless the amount of mass stripped from each galaxy
approximately preserved the relative ordering of dwarf masses.  We
also measure significant internal metallicity spreads,
$\sigma_{[Fe/H]}$, up to 0.5~dex in several ultra-low luminosity
dwarfs, as listed in Table~\ref{masstable}.  This suggests that stars
formed in multiple star formation episodes, rather than a single
burst, and firmly distinguishes these faint dwarfs from globular
clusters, which do not contain mixed stellar populations.  This is
clearly the case for Leo~T, which shows evidence for multiple stellar
populations from its color-magnitude diagram \citep{irwin07}.
However, a metallicity spread is the only evidence of multiple star
formation episodes in the other dwarfs.  Further investigation into
the detailed abundances of these will provide a much clearer picture
of star formation in these low mass objects.

\subsection{Comments on Individual Galaxies}
\label{comments}
\begin{list}{$\bullet$}{\leftmargin=0.0in \rightmargin=0.0in
\topsep=-0.0in \itemsep=0.05in}
\item {\bf Ursa Major II}
\label{uma2}

UMa~II is one of the hardest galaxies to identify based on its
signature in the velocity histogram (see Figure \ref{uma2_obsplot}{\it
  c}), but a clear peak at $-117$~\kms\ emerges once the foreground
dwarf stars are screened out by their \ion{Na}{1} equivalent widths.
Our measured velocity and velocity dispersion are in good agreement
with those of \citet{martin07}.  We identify 20 member stars in UMa~II
out of 236 targeted sources, which represents our lowest detection
rate for any of the galaxies.  However, this is at least partly a
result of our attempt to focus on stars in the outlying clumps noted
by \citet{zucker06b} rather than the main body of the dwarf.  There
are two additional stars we classify as non-members that could in fact
be associated with UMa~II.  One of these stars has a velocity of
$-95.6 \pm 2.3$~\kms, just over 3~$\sigma$ away from the systemic
velocity; including this star as a member would increase the velocity
dispersion of UMa~II to $8.2 \pm 1.6$~\kms.  The other candidate
member, SDSSJ084947.6+630830, was observed on both the first and third
nights of our run and shows a velocity shift of 52.5~\kms\ while also
changing spectral type from a K giant to a horizontal branch star.  We
suspect that this star is an RR Lyrae variable, and its mean velocity
and apparent magnitude suggest that it is plausibly associated with
UMa~II, but we must exclude it from our velocity dispersion
calculation because of its large velocity variability.  Future
observations of this star could provide improved constraints on the
distance of UMa~II.

UMa~II is a clear outlier from the $M_{V}-\sigma$ trend defined by the
other galaxies in Figure \ref{sigmaplot}\emph{a}, with a dispersion of
6.7~\kms\ despite its incredibly low luminosity.  The irregular
appearance, proximity to the Milky Way, and low luminosity of UMa~II
led \citet{zucker06b} to suggest in their discovery paper that this
galaxy might be in the process of tidal disruption.  \citet{belok06b}
and \citet{fellhauer07} have argued that UMa~II is the progenitor of
the recently discovered stellar tidal stream known as the Orphan
Stream.  In \S\,\ref{tidal}, we add additional kinematic and abundance
evidence supporting the hypothesis that UMa~II is a tidally disrupting
satellite that may be associated with the Orphan Stream.

\item {\bf Leo T}

Leo~T is unique among the new dwarfs in that it contains gas
($M_{\rm HI}/M_{\rm star} \sim 1$) and has formed stars in the relatively
recent past \citep{irwin07}.  We measure a mean velocity of $38.1 \pm
2.0$~\kms\ and a stellar velocity dispersion of $7.5 \pm 1.6$~\kms, in
excellent agreement with the \hi\ velocity and gas velocity dispersion
measured by Ryan-Weber et al. (in preparation).  We do not detect any
evidence for a cold stellar population to match the cold gas component
at the center of the galaxy, but our sample of 19 member stars is not
large enough for a significant detection of such a component.  One
would also only expect the youngest blue stars, which are not sampled
by our observations, to have kinematics similar to the cold gas.
Leo~T is now one of very few dwarf galaxies that have well-measured
kinematics from both the stars and the gas, and the agreement between
the two indicates that the gas is accurately tracing the gravitational
potential of the galaxy.  In such a small system, many other effects
could contribute to the velocity dispersion of the gas, but those
contributions appear not to be significant.  If this result also
applies to other dwarfs then \hi\ kinematics can be used to measure
their masses reliably, which is useful because in many cases the gas
extends to larger radii than the stars do.

\item {\bf Ursa Major I}

\citet{kleyna05} reported Keck/HIRES spectra of 5 UMa~I member stars,
obtaining a systemic velocity of $-52.45 \pm 4.27$~\kms\ and a
velocity dispersion of $9.3^{+11.7}_{-1.2}$~\kms.  We reobserved all 7
of the stars from the \citeauthor{kleyna05} sample (including the 2
non-members), and find excellent agreement on individual velocity
measurements as discussed in \S\,\ref{measure_v}.  With our larger
sample of 39 member stars, our mean velocity for UMa~I is in good
agreement with that of \citeauthor{kleyna05}, but our dispersion is
somewhat lower than they measure (the disagreement is at less than
95\% confidence).  We rule out the extremely high velocity dispersions
of up to $\sim20$~\kms\ allowed by \citeauthor{kleyna05}'s data.
Using the same luminosity for UMa~I that \citeauthor{kleyna05} assumed
($M_{V} = -6.75$ from \citealt{willman05a}), we naturally find a
somewhat lower mass-to-light ratio ($355\pm220$~\msun/\lsun) than they
calculate, but with the revised magnitude of $M_{V} = -5.5$ measured
by \citet{belok06}, M/L becomes significantly larger.  Our derived
velocity dispersion is significantly lower than that obtained by
\citet{martin07}, which may indicate that their uncertainties have
been underestimated (see \S \ref{measure_v}).  We do not detect any
evidence for the kinematically cold component ($\sigma < 3.4$~\kms)
suggested by \citeauthor{martin07}, despite a sample of stars that is
a factor of $\sim2$ larger.

UMa~I lacks a published distance uncertainty.  \citet{willman05a}
estimated a distance of 100~kpc from comparisons with the CMD of
Sextans and theoretical isochrones.  We use $\chi^{2}$ fits of the M92
RGB and M13 HB fiducial tracks to our sample of radial velocity member
stars to measure a more accurate distance modulus for UMa~I of
$20.13^{+0.18}_{-0.17}$~mag, corresponding to a distance of
$106^{+9}_{-8}$~kpc.  We increase the assumed absolute magnitude of
UMa~I to $M_{V} = -5.6$ to compensate for this slight increase in
distance.

\item {\bf Leo IV}

With only one slitmask devoted to it, and a total exposure time of
less than an hour, Leo~IV is the least-well studied galaxy in our
sample.  It also appears to have the smallest velocity dispersion,
although with only 18 member stars and larger-than-average
uncertainties on many of them, the uncertainty on the dispersion is
significant.  The dispersion of Leo~IV also depends critically on our
assumptions about membership.  Two candidate member stars that we have
rejected would significantly influence its properties if they were
included.  One of these stars, although it is located just above blue
end of the horizontal branch, has a velocity of $v = 160.1 \pm
5.9$~\kms\ that is well beyond the 3~$\sigma$ velocity range for the
galaxy.  Adding this star as a member would dramatically increase the
velocity dispersion to $6.4 \pm 2.0$\,\kms.  We therefore reject this
star as a non-member (or possible binary system).  The second star
presents a more ambiguous case.  It has a velocity of $120.2 \pm
2.8$~\kms, which is consistent with membership.  It is located outside
the half-light radius, but is close to 2 other member stars.  However,
inspection of the spectrum reveals that the Ca triplet lines for this
star appear to be double-peaked, and that a more appropriate velocity
for this star may be $\sim129$~\kms.  Given that this star may be a
binary and that its true velocity is uncertain, we consider the safest
approach to be removing it from the sample.  If the star is included
with a velocity of 120.2~\kms\ the dispersion of Leo~IV would be $5.0
\pm 1.4$~\kms; if the star's velocity is 129~\kms\ the effect on the
velocity dispersion is negligible.

\item {\bf Coma Berenices}

Coma Berenices has the lowest luminosity of the new Milky Way
satellites, and is located firmly in the low velocity dispersion half
of the sample.  One of the 59 assumed member stars has a velocity just
outside the 3~$\sigma$ limit, at $v = 83.1 \pm 2.8$~\kms.  This star
is located $\sim0.3$~mag away from the blue edge of the subgiant
branch, and could be an evolved blue straggler in Com.  On the other
hand, if we reject this star from the sample as a foreground
contaminant, the velocity dispersion of Com declines to $3.8 \pm
0.8$~\kms.  Com is unique among the ultra-faint dwarfs in that its
member stars span a wide range of $g-i$ colors near the main sequence
turnoff (MSTO), although only UMa~II is nearby enough to detect such
stars among the other galaxies.  The photometric uncertainties are not
large enough to account for this spread.  The stars on the blue side
of the MSTO could be blue stragglers, but the presence of a few
similarly situated stars on the red side suggests that we might
instead be seeing the effects of multiple stellar populations with
different ages (and hence MSTO luminosities and colors) in Com.
Although it appears in Figure \ref{coma_obsplot}{\it c} that there may
be velocity substructure in Com, this is partly a result of the chosen
velocity binning, and a two-sided Kolmogorov-Smirnov (KS) test
indicates that the observed velocity distribution is consistent with a
Gaussian at the 17\% confidence level.

In terms of its luminosity, radius, and proximity to the Milky Way,
Com is most similar to UMa~II among all of the ultra-faint satellites.
Accordingly, we consider the possibility that, like UMa~II, Com is in
the process of being tidally disrupted.  We discuss the available
evidence for and against disruption n \S\,\ref{tidal}, but conclude
that Com is more likely to be a bound, dark matter-dominated object.

\item {\bf Canes Venatici II}

CVn~II is a very faint and very compact dwarf with a low mass.  We
have rejected one star as a member that is located well outside the
half-light radius and has a velocity that is almost 4~$\sigma$ away
from the mean velocity.  If we instead opt to include this star, the
velocity dispersion of the galaxy increases somewhat to $5.1 \pm
1.1$~\kms.

\item {\bf Canes Venatici I}

Along with Leo~T, CVn~I is the only other of the new dwarfs to display
a broad red giant branch, indicating a significant spread in
metallicity and/or age among its stellar population.  For Leo~T, such
a result is not surprising because young stars are visible in its CMD
\citep{irwin07} and it still contains gas, but it is somewhat less
expected in a tiny, gas-poor system like CVn~I.  It is not currently
understood how low-mass dSphs managed to hold on to enough gas to form
stars over an extended period of time (although see, e.g.,
\citealt{rg05,marcolini06}), but the same phenomenon is observed in
all of the more luminous Milky Way dSph satellites.  In this sense,
CVn~I may have more in common with the previously-known dSphs (it
approaches the lower bound of their luminosities) than it does with
its ultra-faint SDSS cousins (which have much lower luminosities).
While other dwarfs in our sample show evidence for metallicity
spreads, CVn~I and Leo~T --- the two brightest of the ultra-faint
dwarfs, and two of the three most massive (along with UMa~I) --- are
the only two with RGB spreads as well.  This result suggests that the
critical mass and luminosity for dwarf galaxies to maintain the
ability to form multiple stellar populations are $M_{V} \simless -7$
and $M \simgtr 10^{7}$~\msun\ (although see \S\,\ref{metals}).

\citet{ibata06} measured the velocities of 44 stars in CVn~I and
reported two distinct kinematical components to the galaxy: a
centrally concentrated metal-rich population with a velocity
dispersion of less than 1.9~\kms\ at 99\% significance and a more
extended metal-poor component with a velocity dispersion of
$13.9^{+3.2}_{-2.5}$~\kms.  Our much larger sample of 214 CVn~I member
stars does not reveal any trace of this dichotomy.  Dividing our
sample in half using the same metal-rich/metal-poor cutoff as
\citeauthor{ibata06} ([Fe/H]$ = -2.0$), we find $\sigma = 8.1 \pm
0.8$~\kms\ for the metal-rich stars and $\sigma = 7.2 \pm
0.5$~\kms\ for the metal-poor stars.  Even limiting the metal-rich
selection to the most metal-rich $\sim10$\% of the stars (in case the
population detected by \citealt{ibata06} represents only a small
fraction of the overall stellar population) does not reveal any
evidence for a cold component.  We also do not detect any tendency for
the metal-rich stars to be more centrally concentrated in the galaxy
than the metal-poor stars.  Finally, we measure a mean velocity of
$30.9 \pm 0.6$~\kms, inconsistent with the value of
$\sim24.5$~\kms\ determined by \citet{ibata06}.  Because of our very
large sample of member stars and our repeat measurements of individual
stars to constrain our errors carefully, we conclude that there is no
kinematically cold population in CVn~I, and there is no detectable
difference between the kinematics and spatial distributions of the
metal-rich and metal-poor stars.

\item {\bf Hercules}

The only galaxy in which we detect possible evidence of kinematic
substructure is Hercules.  There is a peak in the velocity histogram
containing 9 stars between 41 and 43~\kms\ (compared to the mean
velocity of $45.0 \pm 1.1$~\kms), with the remaining 21 stars
distributed more broadly between 30 and 60~\kms.  Given an
intrinsically Gaussian distribution with the same mean velocity and
velocity dispersion that we measure for Hercules, the likelihood of
finding as many as 9 stars out of 30 in such a narrow, offset peak is
only $\sim1$\%.  However, a two-sided KS test indicates that the
observed velocity distribution of stars in Hercules is consistent with
a Gaussian at the 43\% level.  We conclude that there is not yet
statistically significant evidence of velocity substructure in
Hercules.  If the velocity substructure does turn out to be real, it
could be a sign in favor of the tidal disruption hypothesis advanced
by \citet{coleman07}.

Hercules contains two stars whose membership status is difficult to
determine.  The first of these, which is excluded from our member
sample, has a velocity exactly on the mean velocity of the galaxy, but
is offset almost 0.3~mag from the RGB and is therefore rejected on the
basis of the photometry.  Because its velocity is so close to the mean
of the other stars, this star would have a negligible effect on the
measured velocity dispersion.  The other candidate star, which does
satisfy our membership criteria, has a velocity that is just within
3~$\sigma$ of the mean velocity, at $30.7 \pm 2.2$~\kms.  The Na
equivalent width and metallicity of this star are also near the edge
of the membership ranges; if we remove this star from the sample then
the velocity dispersion of the galaxy would decrease to $4.2 \pm
0.9$~\kms.

\end{list}

\subsection{Tidal Disruption}
\label{tidal}

Two of the dwarfs presented in this paper show at least some evidence
for on-going tidal disruption by the Milky Way.  As mentioned in \S
\ref{comments}, the properties of UMa~II and perhaps Com appear to be
affected by these interactions.

UMa~II is located very close to the Milky Way, second only to
Sagittarius (which is the archetype of tidally disrupting dwarfs)
among the known dSphs.  \citet{zucker06b} noted that UMa~II appears
irregular and its stars are broken up into several subclumps.
\citet{belok06b} pointed out that the Orphan Stream lies along a great
circle intersecting the position of UMa~II, and our measured radial
velocity of $-116.5 \pm 1.9$\,\kms\ is in reasonable agreement with
the $100$\,\kms\ predicted by \citet{fellhauer07} if UMa~II is
associated with the Orphan Stream.  \citeauthor{fellhauer07}~also
predict a roughly north-south velocity gradient over several degrees
within UMa~II.  Although our member sample only spans a declination
range of 13.6\arcmin, we do detect a modest correlation between radial
velocity and declination among the member star (correlation
coefficient of $-0.40$), in the same sense as predicted.  More
significantly, we find strong evidence for a difference in the mean
velocity between the eastern and western halves of the galaxy, with
the stars on the eastern side having a velocity $8.4 \pm
1.4$~\kms\ larger than those on the western side.  It is highly
unlikely that a galaxy as small as UMa~II would show significant
coherent rotation, so this velocity gradient strongly suggests that
UMa~II is distorted by tidal forces.  As noted previously, UMa~II is
also a clear outlier from the $M_{V}-\sigma$ trend shown in Figure
\ref{sigmaplot}\emph{a}.  This galaxy therefore either has a
mass-to-light ratio several times larger than any other dwarf
(Table~\ref{masstable}), or its velocity dispersion has been inflated
by the tidal field of the Milky Way.  Finally, UMa~II has a
metallicity $\simgtr0.5$~dex higher than would be expected from the
luminosity-metallicity relationship shown in Figure~\ref{feh_fig}.
Its metallicity is more appropriate for a system with $M_{V} \approx
-10$ (250 times more luminous than UMa~II).  Taken together, all of
these independent results make a strong case for the imminent tidal
disruption of UMa~II, and we are not aware of any observational
evidence suggesting that UMa~II is bound.

Coma Berenices presents an intriguing counterpoint to UMa~II.  It
shares some notable properties with UMa~II, including an exceptionally
low luminosity ($M_{V} = -3.7$, compared to $M_{V} = -3.8$), a
location near the Milky Way (44~kpc instead of 32~kpc), and an
unexpectedly high stellar metallicity.  As with UMa~II, we find a
modest correlation of velocity with position in the galaxy
(correlation coefficient of velocity with right ascension = $-0.24$).
Dividing the galaxy in half along the minor axis, we find a mean
velocity of $93.3 \pm 1.1$~\kms\ for the northwestern side and a mean
velocity of $98.8 \pm 0.5$~\kms\ for the southeastern side.  This
velocity difference is significant at the 4~$\sigma$ level.  As with
UMa~II, it is not expected that galaxies of this size are rotationally
supported, so if this velocity gradient is real it suggests that Coma
Berenices, like UMa~II, may be distorted by tidal forces.  On the
other hand, there are no known tidal streams that are plausibly
associated with Com, its velocity dispersion is approximately what
would be expected given its luminosity, and its stellar distribution
is not noticeably more irregular than those of the other ultra-faint
dwarfs (there are two bright stars immediately to the north of Com
that may be responsible for the apparent distortion of the isopleths
in that direction pointed out by \citealt{belok07}).  We also note
that, with a smaller half-light radius (and larger central density;
see \S \ref{density}) than any other Local Group dwarf, Com may be
more robust to disruption than some of its counterparts.  While the
available evidence is suggestive of the possibility that Coma
Berenices could be tidally disrupting, the situation is not nearly as
clear-cut as it is for UMa~II.  We therefore treat Com as a bound,
dark matter-dominated object for now, while recognizing that future
observations (most importantly, identification of an associated
stellar stream) could change this picture.

For the other six galaxies in our sample, we do not detect any
statistically significant velocity gradients or other evidence
suggesting tidal disruption.

\section{DISCUSSION}
\label{discussion}

\subsection{The Missing Satellite Problem}
\label{missingsats}

Understanding the nature of the ultra-faint dwarf galaxies and
determining their impact on the missing satellite problem is one of
the key goals of this work.  Our observations show that with the
likely exception of UMa~II (and possibly Coma Berenices as well) the
ultra-faint dwarfs seem to be dark matter-dominated systems, with
masses lower than those of the previously-known dSphs and very large
mass-to-light ratios.  These galaxies are currently the darkest known
stellar systems in the universe.

Determining the importance of the effect that the new dwarfs have on
the abundance of satellite galaxies around the Milky Way requires
having a way to compare observed galaxy properties to the properties
of subhalos in N-body simulations.  The simplest possible approach is
to estimate the halo circular velocities of the ultra-faint dwarfs as
$v_{\rm circ} = \sqrt{3} \sigma$ \citep{klypin99}, assuming that the
observed dispersions are equivalent to the maximum dispersions reached
in each galaxy, and that the stars have negligible orbital anisotropy.
Although these assumptions may not be correct in detail, if we use
cumulative satellite distributions then the results of this exercise
are relatively insensitive to them.  The circular velocities of dark
matter subhalos in the simulations can be measured robustly, giving us
an appropriate point of comparison.  We note that a more accurate
means of comparing observed dwarfs to simulated subhalos is to use the
mass contained within 0.6~kpc, which is better constrained by the
observations than the halo circular velocity is \citep{strigari07b};
these calculations will be presented in a future paper (Strigari et
al., in preparation).  Using the above approximation, we find that the
ultra-faint dwarfs have circular velocities from $v_{\rm circ} =
6-13$~\kms\ (for plotting and comparison purposes, we round the
circular velocity of Leo~IV up to 6~\kms).  Because the fifth data
release of the SDSS, where all of the new Milky Way satellites have
been discovered, only covers 8000 deg$^{2}$ of sky, we must weight
each of the new dwarfs by a factor of $\sim5$ to account for the
additional ultra-faint dwarfs likely to be discovered once the rest of
the sky has been similarly surveyed.

\begin{figure}[t!]
\epsscale{1.24}
\plotone{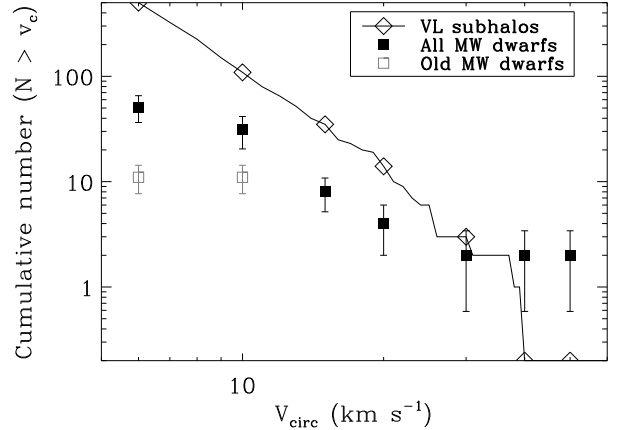}
\caption{Cumulative number of Milky Way satellite galaxies as a
  function of halo circular velocity.  The filled black squares
  include the new circular velocity estimates from this paper (plus
  Bo{\" o}tes, but excluding UMa~II) as well as all of the previously
  known Milky Way dwarfs.  The open gray squares show the observed
  distribution without the new ultra-faint dwarfs.  We assume Poisson
  errors on the number count of satellites in each bin (computed
  independently for the new and old dwarfs), although the true
  uncertainties may be larger.  The solid line plus diamonds
  represents the subhalo abundance within the virial radius of the Via
  Lactea N-body simulation \citep{diemand06}.}
\label{substructure_plot}
\end{figure}

We display the cumulative number of Milky Way satellites as a function
of circular velocity in Figure \ref{substructure_plot}.  We assume
Poisson uncertainties on the total number of dwarfs ($\mbox{dN} =
\sqrt{\mbox{N}_{\rm old} + 5^{2}\mbox{N}_{\rm new}}$, where $N_{\rm
  old}$ and $N_{\rm new}$ refer to the previously-known and
newly-discovered dwarf galaxies, respectively).  For comparison, we
include the subhalo circular velocity function from the recent Via
Lactea simulation, currently the highest resolution (234 million
particles) N-body simulation of a Milky Way-size galaxy
\citep*{diemand06,diemand07}.  This simulation assumes the best-fit
WMAP three-year cosmological parameters: $\Omega_{m} = 0.238$,
$\Omega_{\Lambda} = 0.762$, $h = 0.73$, $n = 0.951$, and $\sigma_{8} =
0.74$ \citep{spergel07}.  The Via Lactea subhalos include all bound
halos located within the virial radius (389~kpc) of the main halo (see
also \S \ref{completeness}).  The addition of the new dwarfs, combined
with the correction for the sky area that has yet to be observed with
sufficient sensitivity, substantially changes the appearance of the
substructure problem.  The previously-known Milky Way satellite
galaxies have a nearly flat circular velocity function below $v_{\rm
  circ} = 15$~\kms, causing a discrepancy with the predictions that
worsens with decreasing mass and reaches well over an order of
magnitude below $v_{\rm circ} = 10$~\kms.  With the ultra-faint dwarfs
included we now see a rising circular velocity function and a
satellite underabundance of a factor of $\sim4$ for halos with masses
between $v_{\rm circ} = 10$~\kms\ and 20~\kms.  At $v_{\rm circ} =
6$~\kms\ the discrepancy increases again towards an order of
magnitude, but if the current observational census is still incomplete
at the faint end, this is the mass range where that would manifest
itself.  The ultra-faint dwarfs significantly fill in the gap for
satellites in the two lowest mass bins, but have masses that are too
small to affect the satellite deficit at higher circular velocities.

\subsubsection{Proposed Solutions to the Missing Satellite Problem}
\label{solutions}

\begin{figure}[t!]
\epsscale{1.24}
\plotone{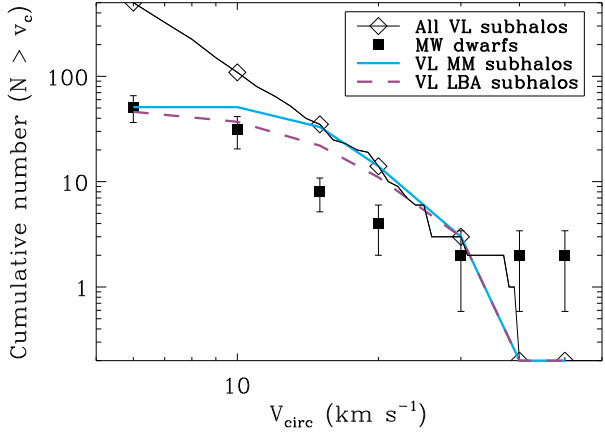}
\caption{Outcome of two proposed solutions to the missing satellite
  problem.  As in Figure \ref{substructure_plot}, the filled black
  squares include the new circular velocity estimates from this paper
  (plus Bo{\" o}tes, but excluding UMa~II) as well as all of the
  previously known Milky Way dwarfs, and the solid line plus diamonds
  represents the subhalo abundance within the virial radius of the Via
  Lactea N-body simulation \citep{diemand06}.  The solid cyan curve
  shows the circular velocity distribution for the 51 most massive Via
  Lactea subhalos at $z=0$.  The dashed purple curve illustrates the
  circular velocity distribution for the 51 Via Lactea subhalos that
  had the largest masses at the time they were accreted by the main
  halo.}
\label{substruc_solns}
\end{figure}

\begin{figure}[t!]
\epsscale{1.24}
\plotone{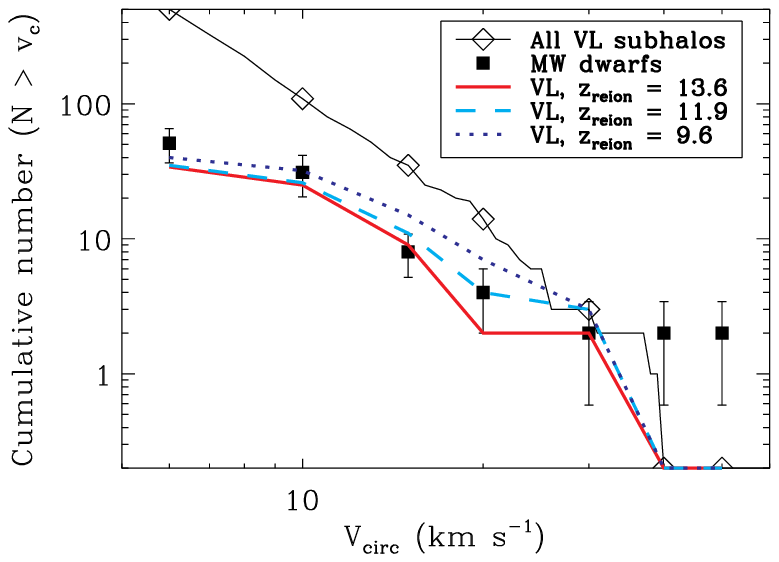}
\caption{Effect of reionization on the missing satellite problem.  As
  in Figure \ref{substructure_plot}, the filled black squares include
  the new circular velocity estimates from this paper (plus Bo{\"
    o}tes, but excluding UMa~II) as well as all of the previously
  known Milky Way dwarfs, and the solid line plus diamonds represents
  the subhalo abundance within the virial radius of the Via Lactea
  N-body simulation \citep{diemand06}.  The solid red curve shows the
  circular velocity distribution for the 51 most massive Via Lactea
  subhalos at $z=13.6$, the dashed cyan curve at $z=11.9$, and the
  dotted blue curve at $z=9.6$.}
\label{substruc_reion}
\end{figure}

Using these new data, we can test a number of proposed astrophysical
solutions to the missing satellite problem.  For example, the observed
dwarf galaxies could inhabit the most massive subhalos at the present
day \citep{stoehr02}, the subhalos that collapsed at the highest
redshift \citep{bkw00}, or the subhalos that were the most massive at
the time they were accreted by the Milky Way \citep{kravtsov04}.  We
show the results of these tests in Figures \ref{substruc_solns} and
\ref{substruc_reion}.  To compare the observed dwarfs to the most
massive (MM) subhalos, we identified the 51 halos (to match the number
of Milky Way satellites projected to be found once the remainder of
the sky has been surveyed) located within the virial radius that have
the largest total masses at the present day in the Via Lactea
simulation.  The circular velocity function of these subhalos is
plotted as the solid cyan curve in Figure \ref{substruc_solns}.  Note
that because we chose the total number of subhalos to match the total
number of Milky Way dwarfs, the agreement between the observed
distribution and the cyan curve in the lowest mass bin is trivial.
Another possibility is to compare the observed circular velocity
function with the circular velocity function of the subhalos that were
most massive when they were accreted (dashed purple curve in Figure
\ref{substruc_solns}).  We selected the largest before accretion (LBA)
subhalos from the Via Lactea simulation as the halos located within
the virial radius of the main halo at $z = 0$ that had the largest
circular velocities at any point in the past.  Again, the agreement at
the low-mass end is simply a result of our choice of the top 51
subhalos from the simulation.\footnote{The largest-before-accretion
  subsample at the present day (dashed purple curve in Figure
  \ref{substruc_solns}) actually only has 46 objects with $v_{\rm
    circ} > 6$~\kms\ because 5 of the subhalos lost so much mass by
  $z=0$ that they end up with even lower present-day circular
  velocities than are shown in the plot.}  If the observed dwarf
galaxies inhabit only the most massive dark matter subhalos around the
Milky Way, the shape of the mass function of the most massive subhalos
fails to match the shape of the observed mass function.  Using the
subhalos that were most massive at the time they were accreted instead
of the ones most massive today (i.e., allowing for mass lost by tidal
stripping) brings the subhalo mass function slightly closer to the
observed one, but there are still a factor of $\sim3$ too few dwarfs
in the $v_{\rm circ} = 10-30$~\kms\ range.

The final astrophysical solution we consider is that only halos that
collapsed prior to reionization were able to form significant numbers
of stars \citep[e.g.,][]{bkw00,somerville02,moore06}.  Among the Via
Lactea subhalos that are located within the virial radius at $z=0$, we
select the objects with the 51 largest values of $v_{\rm circ}$ at
various high redshifts.  The results of this test are displayed in
Figure \ref{substruc_reion}.  The solid red, dashed cyan, and dotted
blue curves represent the subhalos that would be selected if $z_{\rm
  reion} = 13.6$, 11.9, and 9.6, respectively.  \emph{If reionization
  occurred around redshift $9-14$, and dwarf galaxy formation was
  strongly suppressed thereafter, the circular velocity function of
  Milky Way satellite galaxies approximately matches that of CDM
  subhalos.}  If reionization occurred at $z \simless 8$, we again
find an underabundance of Milky Way dwarfs with $v_{circ} =
15-30$~\kms\ compared to theoretical models, although we note that the
individual subhalo $v_{\rm circ}(z)$ histories in the Via Lactea
simulation are noisy at high redshift, and the number of objects in
these bins is relatively low.  We therefore suggest that the observed
mass function of Milky Way satellite galaxies constrains reionization
to have taken place before $z=8$, in agreement with the 3-year WMAP
results from measurements of the cosmic microwave background
\citep[$z_{\rm reion} = 10.9^{+2.7}_{-2.3}$;][]{page06}.  However,
there are a number of caveats to this analysis: (1) the extrapolation
of dwarf galaxy abundances from the SDSS DR5 sky coverage to the whole
sky must be reasonable, (2) the observed velocity dispersions must
provide a reasonable estimate of the halo circular velocities, (3) the
primary physical mechanism responsible for suppressing the formation
of galaxies in low-mass dark matter halos must be reionization, (4)
the cosmology used for the Via Lactea simulation
\citep{diemand06,diemand07} --- particularly the low value of
$\sigma_{8}$ --- must be a good match to the cosmology of our
universe, and (5) the main halo simulated in Via Lactea must be a
reasonable representation of the Milky Way.  We also note that while
WMAP and most other observational probes are sensitive to the mean
reionization history of the universe, the dwarf galaxies observed in
the study are sensitive primarily to the reionization history \emph{of
  the Local Group}.  If reionization was indeed responsible for the
low abundance of Galactic satellites, then the Milky Way and/or M31
must have been undergoing vigorous enough star formation to ionize the
intergalactic medium of the Local Group at $z > 8$.

\subsubsection{Observational Incompleteness and the Comparison Radius
  in Simulations}
\label{completeness}

One of the important assumptions involved in our analysis in \S\S
\ref{missingsats} and \ref{solutions} is the choice of the radius in
the simulations out to which satellites should be counted.  In the
ideal case, this radius should be the virial radius, as we have used,
but in reality the comparison between observations and simulations is
only meaningful in the regime where the observations are complete.

The observational census for Milky Way satellite galaxies similar to
the brighter dwarf spheroidals ($M_{V} \simless -9$) should be largely
complete by now; the last Milky Way satellite in this luminosity range
to be discovered was Sagittarius \citep*{ibata94}.  Recent searches of
Palomar Sky Survey data, which are sensitive to such galaxies anywhere
within the Local Group, have not detected additional dwarfs
\citep{whiting07,sb02}.  The distribution of Milky Way dwarf galaxies
as a function of Galactic latitude suggests that additional relatively
bright dwarfs remain to be discovered at low latitudes, where
extinction and foreground confusion are serious problems.  The
expected number of such objects if they are distributed uniformly
around the Galaxy is $\sim4$ \citep{mateo98,willman04}, which does not
appear to be enough to significantly affect the missing satellite
problem.

Very recently, \citet{koposov07} have analyzed the detectability of
faint Milky Way satellites in the SDSS DR5 data.  They find that
extremely low-luminosity objects ($M_{V} \simless -5$) may be missed
by SDSS searches if they are located at relatively large distances ($d
\simgtr 100$~kpc), as the horizontal branch and MSTO stars that their
detection relies on become too faint to be reliably detected in the
SDSS.  Galaxies with even lower surface brightnesses than the known
dwarfs ($\mu_{V} \simgtr 30$~\surfb), if they exist, are also likely
to have escaped detection.  However, if there is a correlation between
surface brightness and distance from the host galaxy
\citep[e.g.,][]{mi06} or density and distance (as our data and
\citealt{mayer01b} suggest), there may not be significant numbers of
ultra-faint dwarfs at large distances.  For more luminous dwarfs and
those with higher central surface brightnesses, the current sample of
Milky Way satellites should be reasonably complete.  The
\citet{koposov07} luminosity function of Milky Way satellite galaxies
predicts that there are a total of 57 dwarf galaxies within 280~kpc of
the Milky Way over the whole sky, and a similar number within 420~kpc.
Thus, our much simpler estimates of $46 \pm 14$ dwarf galaxies within
250~kpc and $51 \pm 15$ within 420~kpc from \S \ref{missingsats}
appear to be well-justified.

Incompleteness may still be a significant problem at the extreme faint
end of the luminosity function, as the recent discovery of
Bo{\"o}tes~II reveals \citep{walsh07}.  Our results suggest that
satellites in this luminosity range are not gravitationally-bound
dwarf galaxies.  If these objects are tidally disrupted dwarfs, then
they should still contribute to the Milky Way satellite census, but if
they are simply multiple fragments from larger objects (for example,
if there is a physical connection between Bo{\"o}tes and
Bo{\"o}tes~II) or unusual globular clusters then they do not
correspond to dark matter subhalos in the CDM simulations.  Until
surveys are more complete at faint magnitudes and some kinematic
information is available for this class of objects, their effect on
the missing satellite problem is not clear.

In case there are undiscovered ultra-faint Milky Way dwarfs beyond $d
= 250$~kpc, we repeated our analysis of \S \ref{missingsats} using
subhalos within 250~kpc from the main halo in the Via Lactea
simulation (and discarding Leo~T from the observed sample because it
is beyond this radius).  This smaller radius reduces the overall
number of satellite subhalos by $\sim30$\%, not enough to
significantly change our conclusions.  Limiting the comparison to this
radius shifts the preferred range of reionization redshifts slightly
lower, but also lessens our leverage on determining the redshift of
reionization.

\subsection{The Constant Halo Mass Hypothesis}

\begin{figure*}[t!]
\epsscale{1.20}
\plotone{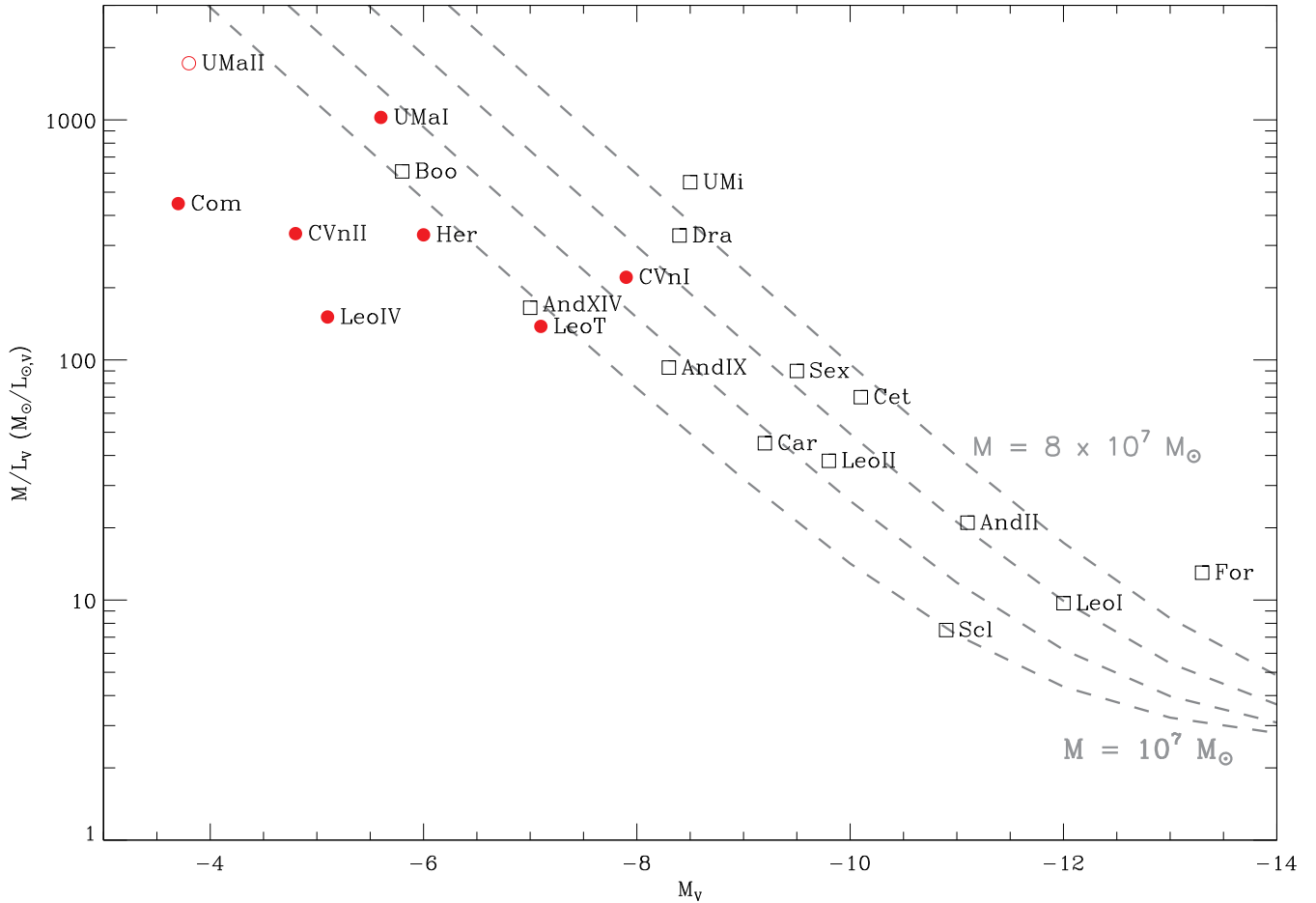}
\caption{Total mass-to-light ratios (in solar units) as a function of
  absolute magnitude for Local Group dwarf spheroidals.  The red
  symbols represent the ultra-faint dwarfs from this paper (including
  Leo~T, which is not really a dSph, and UMa~II, which may be tidally
  disrupted, as an open red circle in the upper left).  The open black
  squares represent all of the dSphs with previously-published
  kinematic data, including satellites of M31 as well as the Milky
  Way.  The dashed gray lines are curves of constant dark matter halo
  mass ($1,2,4,8 \times 10^{7}$~\msun\ from bottom to top), assuming a
  stellar mass-to-light ratio of 2.5~\mlv.  For the previously-known
  Milky Way dwarfs, we recomputed luminosities from \citet{ih95} using
  the most up-to-date distance measurements, and then adjusted the
  mass-to-light ratios from the literature accordingly.  References
  for distance measurements are: Fornax \citep*{saviane00,mg03,gull07},
  Leo~I \citep{bellazzini04}, Sculptor \citep{mateo98}, Leo~II
  \citep*{bellazzini05}, Sextans \citep{lee03}, Carina \citep{d03},
  Ursa Minor \citep{mb99}, and Draco \citep{bonanos04}.  References
  for M/L measurements are: Fornax \citep{walker06}, Leo~I
  \citep{sohn06a,koch07a}, Sculptor \citep{westfall06}, Leo~II
  \citep{koch07b}, Sextans \citep{walker06b}, Carina \citep{munoz06a},
  Ursa Minor \citep{wu07}, Draco \citep*{lokas05}, And~II
  \citep{cote99}, Cetus \citep{lewis07}, And~IX \citep{chapman05}, And
  XIV \citep{majewski07}, and Bo{\" o}tes \citep{munoz06}. }
\label{mateoplot}
\end{figure*}

\citet{mateo93} was the first to point out that observations of the
dSphs known at that time suggested that they were all embedded within
dark matter halos of mass $\sim3 \times 10^{7}$~\msun, independent of
luminosity.  In Figure \ref{mateoplot} we display an updated version
of what has become popularly known as the ``Mateo'' plot, showing the
mass-to-light ratios of all of the Local Group dSphs with measured
kinematics as a function of absolute magnitude.  As seen previously by
\citet{mateo93}, \citet{mateo98}, and \citet{gilmore07}, all of the
galaxies observed prior to this work are approximately consistent with
the picture proposed by \citet{mateo93}.  The results change, however,
when the ultra-faint Milky Way satellites are added.  Although the
brightest of the ultra-faint dwarfs still lie within the same range of
halo mass as their more luminous counterparts, the fainter objects
(Hercules, Leo~IV, CVn~II, and Coma Berenices) are located well below
the extrapolated trend.  These galaxies have much lower halo masses,
and hence their mass-to-light ratios are significantly smaller than
what would be expected if they too were embedded in $\sim3 \times
10^{7}$~\msun\ halos.  Combining the new and old dwarfs, it appears
that there are two distinct regimes: the brighter dwarfs ($M_{V} <
-9$) all have similar mass dark matter halos, but for the fainter
dwarfs ($M_{V} > -9$) M/L saturates at a value of $200-1000$ and the
halo mass declines as luminosity decreases (see Figure
\ref{sigmaplot}\emph{b}).  It therefore appears that the ultra-faint
dwarfs \emph{do not} occupy halos as massive as those of the
``normal'' dSphs; if there is a minimum halo mass for dwarf galaxies,
it is not clear that the observations have yet reached it.

\subsection{Central Dark Matter Densities in the Ultra-Faint Dwarfs}
\label{density}

The observed velocity dispersions and radii of the new dwarfs
constrain their densities as well as their masses and mass-to-light
ratios.  Because these galaxies are highly dark matter-dominated, the
overall densities we derive are essentially equal to the density of
the dark matter halo of each object.  Following \citet{mateo91}, we
can approximate the central density as

\begin{equation}
166 \sigma^{2} \eta/r_{c}^{2},
\end{equation}
where $\eta$ is a numerical parameter that works out to 1 for
plausible density profiles.  The central densities of the new dwarfs
range from $\sim0.08$~\msun~pc$^{-3}$ for CVn~I and Hercules up to
$\sim2.1$~\msun~pc$^{-3}$ for the faintest galaxy, Coma Berenices.
Alternatively, we can calculate mean densities from the total masses
given in Table \ref{masstable} and the core radii.  To compare with
the mean densities of the previously-known Milky Way dSphs tabulated
by \citet{gilmore07}, we assume that the extent of the new dwarfs is
$\sim2$ King core radii.  Again, Coma Berenices is the densest object,
with a mean density of 0.52~\msun~pc$^{-3}$ ($=
20$~GeV/c$^{2}$~cm$^{-3}$), almost a factor of five higher than any of
the previously known dSphs.  This substantially raises the limiting
mass density of $\sim5$~GeV/c$^{2}$~cm$^{-3}$ identified by
\citet{gilmore07} and suggests that there may not be a true physical
ceiling on the densities of dwarf galaxies (as opposed to an
observational ceiling) at all.  However, if Com is in the process of
tidal disruption, as our observations hint, then the highest dark
matter densities in our sample occur in Leo~T and CVn~II and are only
modestly above the density limit of \citet{gilmore07}.

\subsection{Phase-Space Density Constraints from the Ultra-Faint Dwarfs}
\label{phasespace}

\citet{hd00} introduced the parameter $Q \equiv \rho/\sigma^{3}$ as an
estimate of the coarse-grained phase-space density of the dark matter
in galaxy halos.  As discussed by \citet{hd00}, \citet{dh01}, and
\citet{strigari06}, Liouville's theorem implies that observed values
of $Q$ set a hard lower limit on the original phase-space density of
the dark matter.  By finding the systems with the largest observed
values of $Q$, we can therefore constrain the properties of dark
matter and potentially rule out classes of dark matter candidates.
Observations of low-mass spiral galaxies by \citet{simon05} yield
lower limits of $\sim 10^{-6}$~\msun~pc$^{-3}$~(\kms)$^{-3}$ on $Q$
\citep[][Martinez et al., in preparation]{strigari06}, but those are
less restrictive constraints than are provided by the Ly-$\alpha$
forest.  $Q$ values for the ultra-faint dwarfs are listed in Table
\ref{densitytable}.  These values are calculated under the assumption
that the velocity dispersion of the dark matter (which is what $Q$
actually depends on) is equal to the velocity dispersion of the stars.
Our observations show that most of the ultra-faint dwarfs greatly
exceed the phase-space density constraint from the Ly-$\alpha$ forest,
reaching a maximum of $Q = 2.2 \times
10^{-2}$~\msun~pc$^{-3}$~(\kms)$^{-3}$ in Coma Berenices.  Even if the
derived $Q$ of Com has been affected by tidal disruption, all of the
galaxies except UMa~I, CVn~I, and Hercules have $Q$ values $\simgtr
10^{-3}$~\msun~pc$^{-3}$~(\kms)$^{-3}$, two orders of magnitude better
than the Ly-$\alpha$ forest constraint and about an order of magnitude
improvement compared to the previously-known dSphs.  In fact, the dark
matter velocity dispersion is expected to be larger than the stellar
velocity dispersion, so the $Q$ values we derive are upper limits on
the true $Q$ values for these galaxies.  Nevertheless, these $Q$
values will further restrict the allowed parameter space for Warm Dark
Matter particles, and may have an impact on the meta-CDM scenario
proposed by \citet*{skb07}.

\begin{deluxetable*}{l c c c}
\tablewidth{0pt}
\tablecolumns{5}
\tablecaption{Physical and Phase-Space Densities}
\tablehead{
\colhead{Galaxy} & \colhead{$\rho_{0}$}  & 
\colhead{$\bar \rho$} & \colhead{Q} \\
\colhead{} & \colhead{(\msun~pc$^{-3}$)} & 
\colhead{(\msun~pc$^{-3}$)} & \colhead{(\msun~pc$^{-3}$~[\kms]$^{-3}$)}
}
\startdata 
Ursa Major II\tablenotemark{a} & $1.13 \pm 0.60$ & $0.27 \pm 0.18$ & $3.7 \pm 3.1 \times 10^{-3}$ \\
Leo T             & $0.79 \pm 0.36$ & $0.19 \pm 0.10$ & $1.9 \pm 1.5 \times 10^{-3}$ \\
Ursa Major I      & $0.25 \pm 0.08$ & $0.06 \pm 0.02$ & $5.6 \pm 2.9 \times 10^{-4}$ \\
Leo IV            & $0.19 \pm 0.20$ & $0.05 \pm 0.05$ & $5.3 \pm 9.9 \times 10^{-3}$ \\
Coma Berenices    & $2.09 \pm 0.86$ & $0.52 \pm 0.24$ & $2.2 \pm 1.4 \times 10^{-2}$ \\
Canes Venatici II & $0.49 \pm 0.25$ & $0.12 \pm 0.07$ & $5.1 \pm 4.1 \times 10^{-3}$ \\
Canes Venatici I  & $0.08 \pm 0.02$ & $0.02 \pm 0.01$ & $1.7 \pm 0.5 \times 10^{-4}$  \\
Hercules          & $0.10 \pm 0.04$ & $0.02 \pm 0.01$ & $7.7 \pm 5.2 \times 10^{-4}$    
\enddata
\label{densitytable}
\tablenotetext{a}{UMa~II may be a tidally disrupted remnant, which
  would artificially inflate its density.}
\end{deluxetable*}

\section{SUMMARY AND CONCLUSIONS}
\label{conclusions}

We have obtained Keck/DEIMOS spectra of significant samples of stars
in 8 of the new, ultra-faint Milky Way satellite galaxies recently
discovered in the Sloan Digital Sky Survey.  Using a large
spectroscopic data set of radial velocity standard stars observed with
DEIMOS, repeat DEIMOS measurements of stars in dwarf spheroidals and
globular clusters, and DEIMOS and HIRES spectra of the same stars, we
demonstrated that both our velocity measurements and our derived
uncertainties are accurate.  We then measured the velocities of $18-214$
stars in each galaxy, with typical uncertainties of $\sim3.4$~\kms.

From our measurements of individual stellar velocities, we calculated
velocity dispersions for each of the ultra-faint dwarfs.  The velocity
dispersions, which are listed in Table \ref{dispersiontable}, range
from $3.3 \pm 1.7$~\kms\ for Leo~IV up to $7.6 \pm 0.4$~\kms\ for CVn
I, and we showed that the velocity dispersions are correlated with
luminosity (inversely correlated with absolute magnitude).  Under a
set of simple assumptions, we calculated the total masses of the
ultra-faint dwarfs, finding that these objects are the lowest-mass
galaxies currently known.  From the equivalent widths of the Ca
triplet absorption lines we measured the metallicities of the red
giant branch stars in the new dwarfs and derived mean metallicities
ranging from [Fe/H] = $-2.0$ to [Fe/H] = $-2.3$; several of these
galaxies are the most metal-poor stellar systems yet discovered.

We summarize our primary conclusions from this study as follows:

\begin{enumerate}

\item The ultra-faint Milky Way satellites are dark matter-dominated
  dwarf galaxies with lower masses than any other known galaxies.  

\item The only clear exception among the 8 galaxies we observed, as
  well as those previously observed by others, is Ursa Major~II.
  Based on its clumpy appearance \citep{zucker06b}, small
  galactocentric distance \citep{zucker06b}, associated tidal stream
  \citep[][\S \ref{comments}]{fellhauer07}, inflated velocity
  dispersion (\S \ref{sigma}), unusually high metallicity (\S
  \ref{metals}), and possible velocity gradient (\S \ref{comments}),
  we conclude that UMa~II is in the late stages of tidal disruption.
  The other dwarf with $M_{V} \simgtr -4$, Coma Berenices, has a
  similarly high metallicity that indicates it may have suffered
  substantial tidal stripping as well.  Because Com lacks most of the
  other supporting evidence for tidal disruption, we assume for now
  that it is still a bound, dark matter-dominated dwarf, although we
  recognize that future observations may show otherwise.  Based on
  these results, we suggest that $M_{V} \approx -4$ ($3.4 \times
  10^{3}$~\lsun) is the lower limit to the luminosity of
  gravitationally-bound dwarf galaxies.  We therefore predict that
  objects such as Willman 1, Segue 1, and Bo{\" o}tes II will prove to
  be tidally-disrupted remnants.

\item The 6 ultra-faint dwarfs with $M_{V} \simless -4$ follow the
  luminosity-metallicity relationship established by the more luminous
  Local Group dwarfs, and extend the relation by $\sim2$ orders of
  magnitude in luminosity.  The faintest dwarfs, UMa~II and Com, are
  both outliers from this relationship, with metallicities more than
  0.5 dex too large for their luminosities (or conversely,
  luminosities that are more than two orders of magnitude too small
  for their metallicities).  We detect metallicity spreads of up to
  0.5~dex in several objects, suggesting multiple star formation
  epochs.

\item The total mass-to-light ratios of the ultra-faint dwarfs reach
  as high as 1000~\mlv\ (UMa~I).  While the brighter galaxies ($M_{V}
  \simless -9$) have mass-to-light ratios consistent with the
  hypothesis that all dwarf spheroidals are embedded within dark
  matter halos of the same mass, the fainter galaxies depart from this
  trend in the sense that their mass-to-light ratios are too low
  (i.e., they have lower masses).  We therefore suggest that the
  minimum \emph{mass} for dwarf galaxies (as opposed to the minimum
  luminosity mentioned earlier), if there is one, may not have been
  reached yet.

\item The ultra-faint Milky Way satellites, after correcting for the
  sky area not covered by DR5 of the Sloan survey, substantially
  increase the abundance of dwarf galaxies with very low masses
  ($v_{\rm circ} \le 15$~\kms), thereby reducing the satellite deficit
  compared to CDM simulations to a factor of $\sim4$.  Proposals to
  remedy the missing satellite problem by placing the observed dwarf
  galaxies in the most massive dark matter subhalos (at the present
  day) around the Milky Way or in the subhalos that were most massive
  at the time they were accreted by the Milky Way do not reproduce the
  observed shape of the circular velocity function.  If we assume
  instead that only the halos that acquired a significant amount of
  mass ($v_{\rm circ} \simgtr 8$~\kms, varying somewhat with $z_{\rm
    reion}$) before the redshift of reionization were able to form
  stars, then the subhalos from the Via Lactea simulation
  \citep{diemand06,diemand07} approximately match both the total
  number of Milky Way dwarfs and the shape of the circular velocity
  function.

\item The central dark matter densities of the ultra-faint dwarfs are
  as high as 2.1~\msun~pc$^{-3}$ (0.8~\msun~pc$^{-3}$ if Coma
  Berenices is tidally disrupting), significantly larger than those of
  the previously-known dwarf spheroidals.  The phase-space densities
  are also higher than those of other astrophysical systems ($Q >
  10^{-3}$~\msun~pc$^{-3}$~(\kms)$^{-3}$), which will place
  significant limits on non-CDM dark matter models.

\end{enumerate}

\acknowledgements{Data presented herein were obtained at the
  W. M. Keck Observatory, which is operated as a scientific
  partnership among the California Institute of Technology, the
  University of California, and the National Aeronautics and Space
  Administration.  The Observatory was made possible by the generous
  financial support of the W. M. Keck Foundation.  The authors wish to
  recognize and acknowledge the very significant cultural role and
  reverence that the summit of Mauna Kea has always had within the
  indigenous Hawaiian community.  We are most fortunate to have the
  opportunity to conduct observations from this mountain.  JDS
  gratefully acknowledges the support of a Millikan Fellowship
  provided by Caltech and MG acknowledges support from a Plaskett
  Research Fellowship at the Herzberg Institute of Astrophysics of the
  National Research Council of Canada.  We thank George Djorgovski,
  Vasily Belokurov, Leo Blitz, James Bullock, Judy Cohen, Pat C{\^
    o}t{\' e}, J{\" u}rg Diemand, Gerry Gilmore, Raja Guhathakurta,
  Nicolas Martin, Emma Ryan-Weber, Wal Sargent, Peter Stetson, Louie
  Strigari, Beth Willman, and Dan Zucker for helpful conversations,
  and we acknowledge the useful suggestions of the anonymous referee.
  We also thank Michael Cooper and the DEEP2 team for their hard work
  on the DEIMOS data reduction pipeline.  The analysis pipeline used
  to reduce the DEIMOS data was developed at UC Berkeley with support
  from NSF grant AST-0071048.  This research has made use of NASA's
  Astrophysics Data System Bibliographic Services, the NASA/IPAC
  Extragalactic Database (NED) which is operated by the Jet Propulsion
  Laboratory, California Institute of Technology, under contract with
  the National Aeronautics and Space Administration, and the SIMBAD
  database, operated at CDS, Strasbourg, France.}

{\it Facilities:} \facility{Keck:II (DEIMOS)}

\appendix
\section{APPENDIX: COLLECTING UNIFORM DATA FOR THE ULTRA-FAINT DWARF GALAXIES}
\label{data}

The new ultra-faint dwarf galaxies have been discovered by a number of
different authors, which means that their properties have not all been
determined in a uniform manner.  In order to calculate masses and
mass-to-light ratios as consistently as possible for each galaxy, and
to facilitate future studies of these objects, in this appendix we
collect the currently available data on all 12 of the new Milky Way
satellites (see Table~\ref{dwarfdatatable}).

\begin{deluxetable*}{l c c c c c}
\tablewidth{0pt}
\tablecolumns{6}
\tablecaption{Parameters of the Ultra-Faint Milky Way Satellites}
\tablehead{
\colhead{Galaxy} & \colhead{$M_{V}$}  & 
\colhead{distance} & \colhead{$r_{Plummer}$} 
& \colhead{$r_{Plummer}$} & \colhead{References} \\
\colhead{} & \colhead{} & \colhead{(kpc)} &
\colhead{(arcmin)} & \colhead{(pc)} & \colhead{}
}
\startdata 
Ursa Major II     & $-3.8 \pm 0.6 $  & $32^{+5}_{-4}$ & 13.6 & $127 \pm 21$ & (1),(2) \\
Leo T             & $-7.1 \pm 0.3 $  & $417^{+20}_{-19}$ & 1.4  & $170 \pm 15$  & (3) \\
Ursa Major I      & $-5.6 \pm 0.6 $  & $106^{+9}_{-8}$ & 10.0 & $308 \pm 32$ & (2),(4),(5) \\
Leo IV            & $-5.1 \pm 0.6 $  & $158^{+15}_{-14}$ & 3.3  & $152 \pm 17$ & (6) \\
Coma Berenices    & $-3.7 \pm 0.6 $  & $44 \pm 4$      & 5.0  & $64 \pm 7$   & (6) \\
Canes Venatici II & $-4.8 \pm 0.6 $  & $151^{+15}_{-13}$ & 3.0  & $132 \pm 16$ & (6) \\
Canes Venatici I  & $-7.9 \pm 0.5 $  & $224^{+22}_{-20}$ & $8.5 \pm 0.5$ & $554 \pm 63$ & (7) \\
Hercules          & $-6.0 \pm 0.6 $  & $138^{+13}_{-12}$ & 8.0  & $321 \pm 36$ & (6) \\  
\hline
Segue 1           & $-3.0 \pm 0.6 $  & $23 \pm 2$ & 4.5          & $30 \pm 3$ &  (6) \\
Willman 1         & $-2.5 \pm 1.0 $  & $38 \pm 7$ & $1.9 \pm 0.3$ & $21 \pm 5$ &  (8) \\
Bo{\" o}tes II    & $-3.1 \pm 1.1 $  & $60 \pm 6$ & $4.1\pm1.6$  & $72 \pm 28$ &  (9) \\ 
Bo{\" o}tes       & $-5.8 \pm 0.5 $  & $60 \pm 6$ & $13.0\pm0.7$ & $227 \pm 26$ & (3)   
\enddata 

\tablerefs{(1) \citealt{zucker06b}; (2) D. Zucker \& V. Belokurov
  2007, private communication; (3) \citealt{irwin07}; (4)
  \citealt{belok06}; (5) this work; (6) \citealt{belok07}; (7)
  \citealt{zucker06a}; (8) \citealt{willman06}; (9) \citealt{walsh07}
}
\label{dwarfdatatable}
\end{deluxetable*}

With the exception of Willman~1 and Bo{\"o}tes~II, we use the absolute
magnitudes and radii determined by the Cambridge group (Zucker,
Belokurov, Irwin et al.) from SDSS data.  This includes the revised
absolute magnitude of $M_{V} = -5.5$ for UMa~I reported in
\citet{belok06}, which differs substantially from the original value
of $M_{V} = -6.75$ estimated by \citet{willman05a}, although the
uncertainties on both numbers are admittedly large.  Note that our
improved distance for UMa~I of 106~kpc (compared to the previously
reported 100~kpc) requires a corresponding change in the absolute
magnitude to $M_{V} = -5.6$.  The \citeauthor{belok06}~magnitude for
UMa~I is not accompanied by an uncertainty; by analogy to the other
galaxies of similar luminosity we assume an uncertainty of 0.6~mag.
\citeauthor{belok06}~have also re-measured the radius of UMa~I in the
same manner as they did for the other Milky Way galaxies, finding a
Plummer radius of 10\arcmin\ (D. Zucker \& V. Belokurov 2007, private
communication).

UMa~II is described in the discovery paper only as having an angular
extent of $\sim0\fdg5 \times 0\fdg25$ and a half-light radius of
approximately 120~pc \citep{zucker06b}.  Similarly to UMa~I, the
authors have re-fit the light profile using the same method as they
did for the other new dwarfs and measured a half-light radius of
$13\farcm6$ (D. Zucker \& V. Belokurov 2007, private communication).
Note that unlike the Plummer radii we use for the other dwarfs, this
radius is the mean of the Plummer radius and the exponential scale
radius; in most cases the two radii are very similar.

Bo{\"o}tes~II does not have a published distance uncertainty, so given
the angular proximity to Bo{\"o}tes and the apparently identical
distance moduli, we assume the same distance uncertainty for
Bo{\"o}tes~II as \citet{belok06} derived for Bo{\"o}tes.

Most of the new discoveries do not have published uncertainties for
their Plummer radius fits.  For these objects, we assume an
uncertainty of 6\% (the uncertainty given for CVn~I) on the angular
radius for the purposes of calculating the uncertainty on the
corresponding physical radius.

\end{document}